\documentclass[twocolumn,showpacs,amsmath,amssymb]{revtex4}
\input{epsf}

\usepackage{graphicx}
\usepackage{array}
\usepackage{longtable}
\usepackage{rotating,booktabs}
\usepackage{booktabs,threeparttable}
\usepackage{bm}

\begin{document}

\title{Polarizabilities and tune-out wavelengths of the hyperfine ground states of $^{87,85}$Rb}

\author{Xia Wang}
\author{Jun Jiang}
\email {phyjiang@yeah.net }
\author{Lu-You Xie}
\author{Deng-Hong Zhang}
\author{Chen-Zhong Dong}
\email {dongcz@nwnu.edu.cn }

\affiliation{Key Laboratory of Atomic and Molecular Physics and Functional Materials of Gansu Province,
College of Physics and Electronic Engineering, Northwest Normal University, Lanzhou 730070, P. R. China\\
}

\date{\today}

\begin{abstract}
The static and dynamic polarizabilities, and the tune-out wavelengths of the  
ground state of Rb and the hyperfine ground states of $^{87, 85}$Rb 
have been calculated by using relativistic configuration interaction plus 
core polarization(RCICP) approach.
It is found that the first primary tune-out wavelengths of 
the $ 5s_{1/2}, F=1, 2 $ states of $ ^{87}$Rb are 
790.018187(193) nm and 790.032602(193) nm severally, 
where the calculated result for the $ 5s_{1/2}, F=2 $ state 
is in good agreement with the latest high-precision 
measurement 790.032388(32) nm $[\emph{Phys. Rev. A}$ ${\bf 92}$, $052501 (2015)]$.
Similarly, the first primary tune-out wavelengths of the $ 5s_{1/2}, F=2, 3 $ states of $^{85}$Rb are 
790.023515(218) nm and 790.029918(218) nm respectively. Furthermore,  
the tune-out wavelengths for the different magnetic sublevels $ M_{F}$ of each 
hyperfine level $F$ are also determined by considering the contributions of tensor polarizabilities. 
\end{abstract}

 \maketitle

\section{Introduction}

If an atom is placed in an AC electromagnetic field, 
the energy shift due to Stark effect can be written as
\begin{eqnarray}
	\Delta E \approx-\frac{1}{2}\alpha_d(\omega)F^2+ \dots,
\end{eqnarray}
where $ \alpha_d( \omega ) $ is the dynamic dipole polarizability of quantum state at
frequency $ \omega $, and $ F $ is the strength of the
AC electromagnetic field. When the frequency $ \omega $ is zero, 
$ \alpha(0) $ is called static polarizability. 
When the frequency $ \omega $ tends to the certain value, the dynamic 
polarizability goes to zero and the corresponding wavelength 
is called tune-out wavelength. 

With the recent development of atomic manipulation 
and measurement in experimental optical traps, studies on 
polarizabilities of atoms and ions have been of great interest.
The knowledge of static polarizabilities can be
used to evaluate Stark effect \cite{safronova13aa} and 
the blackbody radiation(BBR) shift \cite{safronova13b} 
which are very important to determine the 
uncertainty of atomic clock \cite{porsev06aa,middelmann12a,cheng13aa}.

The tune-out wavelength was initially introduced by LeBlanc and Thywissen \cite{leblanc07a}
and they discussed its application in multispecies atom traps .
The atom trapped in the optical lattice 
is released while the other atoms are still strongly trapped when the wavelength 
of trapping laser is equal to the tune-out wavelength of this atom. 
In addition, high-precision measurement of the tune-out wavelength can be used to 
test atomic structure calculations \cite{holmgren12a}.
Up to now, the tune-out wavelengths of Rb \cite{schmidt16a,leonard15a,catani09a}, 
K \cite{holmgren12a} and metastable states of
He \cite{henson15a} have been measured in experiment.
The longest tune-out wavelength of the ground state
of K is measured with an uncertainty of 1.5 pm \cite{holmgren12a}. 
This experiment provides the most accurate determination of the ratio 
of the $4s-4p_{3/2}$ and $4s-4p_{1/2}$ line strengths of K and the uncertainty
is half as much as the theoretical uncertainty \cite{arora11a}. 
Recently, a tune-out wavelength 
of the $5s_{1/2}, F=2$ state of $^{87}$Rb has been measured 
with an accuracy about 30 fm \cite{leonard15a} by using a condensate interferometer. 
This accuracy is better than the precision of other previous measured tune-out wavelengths 
\cite{lamporesi10a,holmgren12a,herold12a,henson15a}. 
The tune-out wavelength of the $5s_{1/2}, F=1, M_{F}=0 $   
magnetic sublevel of $^{87}$Rb also has been measured 
with sub pm accuracy by Schmidt $\emph{et al.}$ \cite{schmidt16a}.
These experiments give some very good opportunities for testing of the theories.

In this paper, the static and dynamic polarizabilities, and tune-out wavelengths of the 
ground state of Rb and the hyperfine ground states of $^{87, 85}$Rb 
have been calculated by using relativistic configuration interaction plus 
core polarization(RCICP) approach.
Firstly, the wavefunctions, energies, and transition matrix elements 
of fine structure of Rb are computed. 
Then, combining the most accurate $5s-5p_J$ and $5s-6p_J$ matrix elements
\cite{leonard15a,herold12a} with the RCICP results, 
the static and dynamic polarizabilities, and three 
tune-out wavelengths of the $5s_{1/2}$ state are determined.
Finally, after considering the hyperfine splittings, 
the dipole matrix elements between the hyperfine states, 
the static and dynamic polarizabilities, and the tune-out wavelengths 
of the hyperfine ground states of $^{87,85}$Rb are also determined. 
In Sec. II., a brief description of the theoretical method 
is presented. In Sec. III. and Sec. IV., the energies, matrix elements, 
static and dynamic polarizabilities, 
and tune-out wavelengths of the fine and hyperfine structure states are computed. 
In Sec. V., a few conclusions are pointed out.
The unit used in the present calculations is atomic 
unit(a.u.), in which, mass of electron $m_{e}$ and $\hbar$ have the numerical
value 1 and the speed of light is 137.0359991.   

\section{Formulation and calculations}

The RCICP method is used in the present calculations.
The details of calculation method are similar to those reported in \cite{jiang13a,jiang13b}.
The starting point is the Dirac-Fock(DF) calculation for the Rb$^{+}$ ground state.
The single electron orbitals of the core are made up of the linear combinations of some
analytical S-spinors basis functions, which were introduced by Grant and Quiney 
\cite{grant00a,grant07a}. S-spinors can be treated as relativistic 
generalizations of the Slater-type orbitals.

The effective interaction potential of the valence electron with the core is written as
\begin{eqnarray}
	\bm{\emph{H}} = c \bm{\alpha} \cdot \bm{p}+ \beta c^2+V_{core}(\bm{r}),
\end{eqnarray}
where $ \bm{\alpha}$  and $\beta$ are $4 \times 4$ matrices of the Dirac operator, $\bm{p}$ is 
the momentum operator, c is the speed of light \cite{grant07a}.
The core operator is 
\begin{eqnarray}
      V_{core}(\bm{r})=-\frac{Z}{r}+V_{dir}(\bm{r})+V_{exc}(\bm {r})+V_{p}(\bm{r}).
\end{eqnarray}
The direct interaction $ V_{dir}(\bm{r}) $ and exchange interaction $ V_{exc}(\bm{r}) $ 
of the valence electron with the DF core 
are calculated without any approximation. The $ {\ell,j} $-dependent polarization potential
$V_{p}$ is semiempirical and can be written as
\begin{equation}
V_{p}(r) = -\sum_{k=1}^{3} \frac{\alpha_{\mathrm{core}}^{(k)}}{2r^{(2(k+1))}}
             \sum_{\ell,j} g^2_{k,\ell,j}(r) | \ell,j \rangle \langle \ell,j| .
\end{equation}
Here, the factors $\alpha_{\mathrm{core}}^{(k)}$ are the static $k$-th order
polarizabilities of the core electrons.
In the present calculations, dipole polarizability is 9.076 a.u. \cite{johnson83a}, 
quadrupole polarizability is
35.41 a.u. \cite{johnson83a}, and octupole polarizability is 314 a.u. \cite{porsev03a}.
$g^2_{k,\ell,j}(r) = 1-\exp(-r^{(2(k+2))}/\rho^{(2(k+2))}_{\ell,j})$ 
is the cutoff function to make the polarization potential 
finite at the origin. The cutoff parameters
$\rho_{\ell,j}$ that can be tuned to redo the energies of the $ns, np_{J}, nd_{J}$ 
states are listed in Table \ref{tab1}.

\begin{table}
\caption{\label{tab1} The cutoff parameters $ \rho_{\ell,j}$ of the polarization 
potential of Rb$^{+}$.}
\begin{ruledtabular}
\begin {tabular}{ccc}
$ \ell $  &  $J$   & $ \rho_{\ell,j}$(units : a.u.) \\
\hline
$s$    & $1/2$  &  2.4254\\
$p$    & $1/2$  &  2.3448\\
       & $3/2$  &  2.3450\\
$d$    & $3/2$  &  2.8047\\
       & $5/2$  &  2.8222\\
\end{tabular}
\end{ruledtabular}
\end{table}

The effective Hamiltonian of the valence electron is diagonalized in a large
L-spinor basis. L-spinors can be treated as relativistic generalizations
of the Laguerre-type orbitals \cite{grant00a,grant07a}. 
This basis can be enlarged until completeness
without any linear dependence problem.
 
\section{Results of fine structure}

\subsection{Energies}

Table \ref{tab2} gives the present theoretical energy levels 
for a few low-lying excited states of Rb, 
which are compared with experimental energies 
from the National Institute of Science and 
Technology (NIST) tabulation \cite{nistasd15a}. 
The polarization potential parameters $\rho_{\ell,j}$ are 
tuned to give the correct experimental energies of $5s$, 
$5p_{J}$, $4d_{J}$. Hence, the spin-orbit splittings of $5p_{J}$ and $4d_{J}$ are 
the same as experimental values. It is worth noting that  
the spin-orbit splittings of the $6p_{J}$, $7p_{J}$, $5d_{J}$, 
and $6d_{J}$ states are also very close to
experimental values. For example, the spin-orbit 
splittings of $5p_{J}$ and $6p_{J}$ states are $0.0010825$ and $0.0003536$ 
Hartree in theory, which are in good agreement with the 
experimental values $0.0010826$ and $0.0003532$ Hartree. 
The spin-orbit splittings of $4d_{J}$ and 
$5d_{J}$ states are $0.0000023$ and $0.0000131$ Hartree in theory, 
which are also consistent with the experimental 
values $0.0000020$ and $0.0000135$ Hartree.

\begin{table}
\caption{\label{tab2} Theoretical and experimental energy levels (in Hartree)
for a few low-lying excited states of Rb. The energies are given relative
to the energy of the Rb$^+$ core. The experimental data come from
the National Institute of Science and Technology (NIST) tabulation
\cite{nistasd15a}.}
\begin{ruledtabular}
\begin {tabular}{cccc}
State & $J$  & Present      &  Experiment     \\
\hline
$5s$ & 1/2 &  $-$0.1535067  &   $-$0.1535066   \\
$5p$ & 1/2 &  $-$0.0961927  &   $-$0.0961927   \\
     & 3/2 &  $-$0.0951102  &   $-$0.0951101   \\
$4d$ & 5/2 &  $-$0.0653180  &   $-$0.0653178   \\
     & 3/2 &  $-$0.0653157  &   $-$0.0653158   \\
$6s$ & 1/2 &  $-$0.0616926  &   $-$0.0617762   \\
$6p$ & 1/2 &  $-$0.0454285  &   $-$0.0454528   \\
     & 3/2 &  $-$0.0450749  &   $-$0.0450996   \\
$5d$ & 3/2 &  $-$0.0363087  &   $-$0.0364064   \\
     & 5/2 &  $-$0.0362956  &   $-$0.0363929   \\
$7s$ & 1/2 &  $-$0.0335803  &   $-$0.0336229   \\
$4f$ & 7/2 &  $-$0.0314334  &   $-$0.0314329   \\
     & 5/2 &  $-$0.0314333  &   $-$0.0314328   \\
$7p$ & 1/2 &  $-$0.0266661  &   $-$0.0266809   \\
     & 3/2 &  $-$0.0265057  &   $-$0.0265211   \\
$6d$ & 3/2 &  $-$0.0227249  &   $-$0.0227985   \\
     & 5/2 &  $-$0.0227150  &   $-$0.0227881   \\
$8s$ & 1/2 &  $-$0.0211350  &   $-$0.0211596   \\
$5f$ & 7/2 &  $-$0.0201079  &   $-$0.0201073   \\
     & 5/2 &  $-$0.0201077  &   $-$0.0201072   \\
$5g$ & 7/2 &  $-$0.0200232  &   $-$0.0200233   \\
     & 9/2 &  $-$0.0200232  &   $-$0.0200233   \\
\end{tabular}
\end{ruledtabular}
\end{table}

\subsection{Dipole matrix elements}

\begin{table*}
\caption{\label{tab3} Comparison of reduced electric dipole(E1) matrix elements(in a.u.)
for the principal transitions of Rb with experimental results and other theoretical 
calculations.}
\begin{ruledtabular}
\begin {tabular}{cccccc}
Transition               &   RCICP       &  RMBPT all-order \cite{safronova11aa,safronova04a} & RCCSDT \cite{arora12a} & RCCSD \cite{pal07a}  & Expt. \\\hline 
$5s$ - $5p_{1/2}$        & 4.221(21)   & 4.253(34)    & 4.26(3)     &  4.26115   & 4.233(2) \footnotemark[1]  \\
                         &               &              &             &            & 4.2339(16) \cite{leonard15a} \\
$5s$ - $5p_{3/2}$        & 5.962(30)   & 6.003(48)    & 6.02(5)     &  6.01328   & 5.978(4) \footnotemark[1]  \\
                         &             &              &             &            & 5.9760(23) \cite{leonard15a} \\
$5s$ - $6p_{1/2}$        & 0.313(4)    & 0.333        & 0.342(2)    &            & 0.3235(9) \cite{herold12a}  \\
$5s$ - $6p_{3/2}$        & 0.513(5)    & 0.541        & 0.553(3)    &            & 0.5230(8) \cite{herold12a}   \\
$6s$ - $5p_{1/2}$        & 4.150(12)   & 4.145(10)    & 4.1187      & 4.144(3) \\
$6s$ - $5p_{3/2}$        & 6.052(17)   & 6.047(13)    & 6.0145      & 6.048(5) \\
$6s$ - $6p_{1/2}$        & 9.723(17)   & 9.721(24)    & 9.6839      &          \\
$6s$ - $6p_{3/2}$        & 13.660(25)  & 13.647(34)   & 13.5918     &          \\
$4d_{3/2}$ - $5p_{1/2}$  & 8.028(40)   & 8.037(43)    & 7.9802      & 8.07(2)  \\
$4d_{3/2}$ - $5p_{3/2}$  & 3.625(18)   & 3.628(20)    & 3.6029      & 3.65(2)  \\
$4d_{3/2}$ - $6p_{1/2}$  & 5.2257(87)  & 4.717        &             &          \\
$4d_{3/2}$ - $6p_{3/2}$  & 2.2810(40)  & 2.055        &             &          \\
$4d_{5/2}$ - $5p_{3/2}$  & 10.880(54)  & 10.889(58)   & 10.8149     & 10.96(4) \\
$4d_{5/2}$ - $6p_{3/2}$  & 6.846(12)   & 6.184        &             &          \\
$5d_{3/2}$ - $5p_{1/2}$  & 1.297(56)   & 1.616        & 1.184(3)    &          \\
$5d_{3/2}$ - $5p_{3/2}$  & 0.640(26)   & 0.787        & 0.59(2)     &          \\
$5d_{3/2}$ - $6p_{1/2}$  & 18.209(98)  & 18.195(87)   & 18.1341     &          \\
$5d_{3/2}$ - $6p_{3/2}$  & 8.2131(56)  & 8.205(27)    & 8.1778      &          \\
$5d_{5/2}$ - $5p_{3/2}$  & 1.909(77)   & 2.334        & 1.76(3)     &          \\
$5d_{5/2}$ - $6p_{3/2}$  & 24.645(16)  & 24.621(80)   & 24.5410     &          \\
$4d_{3/2}$ - $5f_{5/2}$  & 4.630(96)   & 4.614(39)    & 4.5951      &          \\
$4d_{5/2}$ - $5f_{5/2}$  & 1.238(26)   & 1.234(10)    & 1.2287      &          \\
$4d_{5/2}$ - $5f_{7/2}$  & 5.54(18)    & 5.518(45)    & 5.4948      &          \\
$5d_{3/2}$ - $4f_{5/2}$  & 25.382(11)  & 25.357(56)   & 25.3138     &          \\
$5d_{5/2}$ - $4f_{5/2}$  & 6.786(68)   & 6.779(14)    & 6.7677      &          \\
$5d_{5/2}$ - $4f_{7/2}$  & 30.35(51)   & 30.316(64)   & 30.2657     &          \\ 
$\frac{|\langle5p_{3/2}||D||5s_{1/2}\rangle|^2}{|\langle5p_{1/2}||D||5s_{1/2}\rangle|^2}$ & 1.994(40) & 1.992(65) & 1.997(62) & 1.99145 & 1.995(3) \footnotemark[1] \\
                                                                                          &           &         &             &            & 1.99221(3) \cite{leonard15a} \\
\end{tabular}
\footnotetext[1]{ These values are the average of several experiments \cite{volz96a, simsarian98a, gutterres02a} and 
given by Leonard $\emph{et al.}$ \cite{leonard15a}}
\end{ruledtabular}
\end{table*}

Table \ref{tab3} gives the reduced electric dipole(E1) matrix elements for a number of
low-lying excited states transitions of Rb. The matrix elements are
calculated with a modified transition operator \cite{mitroy10a,lee75a,snider66a},
\begin{eqnarray}
   \bm{r}=\bm{r}-[1-exp(\frac{-r^{6}}{\rho^{6}})]^{1/2}\frac{\alpha_{d}\bm{r}}{r^3},
\end{eqnarray}
The cutoff parameter $ \rho $ used in Eq.(5) is 2.5279 a.u., which is the average of
the $s$, $p$ and $d$ cutoff parameters(note, the weighting of the s is 
doubled to give it the same weighting as the two $p$ and $d$ orbitals). 
The present RCICP calculations are compared with the relativistic
many-body perturbation theory all-order method
(RMBPT all-order) \cite{safronova11aa,safronova04a}
and the relativistic coupled cluster single-double and the important 
valence triple excitation method(RCCSDT) calculations \cite{arora12a}. 
For the $5s-5p_{J}$ transitions, the differences among the present 
RCICP, RMBPT all-order, and RCCSDT theoretical results are not larger than 1\%. 
The present RCICP results have a good agreement with the 
average values of experiments \cite{volz96a, simsarian98a, gutterres02a}
and the results of Leonard $\emph{et al.}$ \cite{leonard15a}.
For the $5s-6p_{J}$ transitions, the present RCICP results  
agree with some available results \cite{herold12a,safronova04a, arora12a} very well,
and the experimental values lie in the middle of present 
RCICP results and other theoretical results.  

The ratio of the line strengths, which are the square of electric dipole matrix elements 
of the $5s-5p_{1/2}$ and $5s-5p_{3/2}$ transitions, is also given 
in Table \ref{tab3}. This ratio should exactly be $2.0$ in the nonrelativistic limit. 
The deviation of this ratio
comes from the slight differences of radial wavefunctions 
for the spin-orbit doublet arising from the small differences of energies \cite{migdalek98a}. 
The present RCICP ratio 1.994(40) is in excellent agreement 
with average experimental value 1.995(3),
larger than the latest experimental ratio 1.99221(3) which has been determined by the 
measurement of tune-out wavelength and the experimental 
matrix element of $5s-5p_{1/2}$ 4.233 \cite{leonard15a}.
So far none of theoretical results are within the latest experimental error bar,
but the RMBPT all-order result is the closest to this latest experimental ratio. 

\subsection{Polarizabilities of the ground state }

The static scalar polarizability is written as
\begin{eqnarray}
 \alpha^{(k)} (0)=\sum_{n} \frac{f^{(k)}_{ni}}{\varepsilon_{ni}^2},
\end{eqnarray}
where $ f^{(k)}_{ni}$ is the oscillator strength, 
and $\varepsilon_{ni}$ is the excitation energy of the transition.
The oscillator strength is defined as
\begin{equation}
	f^{(k)}_{ni} = \frac{ 2|\langle L_iJ_i \| r^{k}C^{k}(r) \| L_nJ_n\rangle 
		|^2 \varepsilon_{ni} } {(2k+1)(2J_n+1)} \ .
\end{equation}

\begin{table}
\caption{\label{tab4} The dipole $\alpha^{(1)} $, quadrupole $\alpha^{(2)} $,
and octupole $\alpha^{(3)}$ polarizabilities(in a.u.) of the $5s_{1/2}$ state of Rb.
}
\begin{ruledtabular} 
\begin {tabular}{lccc}
$5s_{1/2}$              & $\alpha^{(1)}$  & $10^{-3}\alpha^{(2)} $   & $10^{-5}\alpha^{(3)} $  \\\hline
present RCICP           & 317.05(3.10)  &  6.479(1)     & 2.381(44) \\ 
DFCP \cite{tang14a}     & 317.62          & 6.4810        & 2.3783    \\
CICP \cite{mitroy03a}   & 315.7           & 6.480         & 2.378     \\
RCCSD \cite{lim05a}     & 316.17            \\
RCCSD \cite{kaur14a,kaur15a}    & 318.47/318.3(6)  & 6.491(18)  &   \\
MBPT-SD \cite{safronova99a}     & 317.39  \\
RMBPT all-order \cite{derevianko99a, safronova11aa} & 316.4/322(4)    & 6.525(37)    & 2.374(16) \\
RMBPT \cite{porsev03a}          &                  & 6.520(80)      & 2.37 \\
Expt.E$\times$H \cite{molof74a} & 319(6.1)           \\
Expt. \cite{holmgren10a}	& 318.79(1.42)              \\
Expt. \cite{gregoire15a}        & 320.1(6)   \\
\end{tabular}
\end{ruledtabular}
\end{table}

Tabel \ref{tab4} gives the present and some available theoretical and experimental 
dipole, quadrupole and octupole polarizabilities of the $5s_{1/2}$ state of Rb. 
It is found that the present RCICP
results agree with the DFCP results \cite{tang14a} very well. 
The DFCP method is the same as the present RCICP method except that DFCP 
uses the B-spline basis. The RCICP dipole polarizability 
is larger than that calculated by the nonrelativistic configuration interaction 
plus core polarization(CICP) \cite{mitroy03a}, 
RCCSD of Lim $\emph{et al.}$ \cite{lim05a}, and 
the RMBPT all-order method \cite{derevianko99a}, 
but smaller than the RCCSD result of Kaur 
$\emph{et al.}$ \cite{kaur14a,kaur15a}, the earlier 
MBPT-SD result \cite{safronova99a}, and the experimental 
values \cite{molof74a,holmgren10a}.
If the experimental electric dipole matrix elements 
of the $ 5s-5p_{J}$ transitions \cite{leonard15a} 
are used in the calculation of polarizabilities, 
the static dipole polarizability of the $5s$ state is 318.743 a.u., which agrees
with the experimental result \cite{holmgren10a} very well. So the 
differences of static dipole polarizabilities between experiments and 
the present results are mainly from the differences of $ 5s-5p_{J}$ matrix elements.  
The latest experimental value \cite{gregoire15a}, 320.1(6) a.u.,  
is larger than most of the theoretical and other experimental values.

The present quadrupole and octupole polarizabilities 
of the $5s$ state are close to
the results of CICP, the relativistic many-body perturbation theory(RMBPT), 
RCCSD, and RMBPT all-order.
The differences of the present RCICP calculations and other available 
results\cite{mitroy03a,kaur14a,kaur15a,safronova11aa,porsev03a} are not more than 0.6$\%$.

\subsection{Tune-out wavelengths of the ground state}

The dynamic dipole polarizabilities computed with the 
usual oscillator strength sum-rules can be written as  
\begin{eqnarray}
	\alpha^{(1)} (\omega ) =\sum_{n} \frac{f^{(1)}_{ni}}{(\varepsilon_{ni}^2-\omega ^2)},
\end{eqnarray}
The core polarizability is given by a pseudospectral oscillator
strength distribution \cite{mitroy03a}. The distribution is
derived from the single particle energies of Hartree-Fock core
and listed in Table \ref{tab5}. Each separate $(n,l)$ level is 
identified with one transition with a pseudo-oscillator strength
that is equal to the number of electrons in the shell. The excitation
energy is set by adding a constant to the Koopman energies
and tuning the constant until the core polarizability is equal
to the known core polarizability from the oscillator strength sum-rules.  

\begin{table}
\caption{\label{tab5} Pseudospectral oscillator strength distribution for
the Rb$^{+}$. Transition energies $\varepsilon_{n}$ are given in a.u.. }
\begin{ruledtabular} 
\begin {tabular}{lrl}
n  	 &  $\varepsilon_{n}$  & $f_{n}$  \\\hline
1    &  551.524651         & 2.0     \\
2    &   75.117766         & 2.0     \\
3    &   12.201477         & 2.0     \\
4    &    1.592215         & 2.0     \\
5    &   67.974337         & 6.0     \\
6    &    9.575915         & 6.0     \\
7    &    0.878715         & 6.0     \\
8    &    4.800593         & 10.0    \\
\end{tabular}
\end{ruledtabular}
\end{table}

\begin{table}
\caption{\label{tab6} Tune-out wavelengths $\lambda_{zero}$ (in nm) of the
$5s_{1/2}$ state of Rb.}
\begin{ruledtabular} 
\begin {tabular}{cccc}
Transition     & RCICP & RMBPT  & Expt.  \\
\hline
$ 5s-5p_{1/2}$ &       &   \\
               & 790.02765(20)  & 790.0261(7) \cite{leonard15a}& 789.85(1) \cite{catani09a} \\
               &                & 790.034(7) \cite{arora11a}   & 790.018(2) \cite{lamporesi10a} \\
$ 5s-5p_{3/2}$ &       &   \\
               & 423.02428(391) & 423.05(8) \cite{arora11a}    & 423.018(7) \cite{herold12a} \\
$ 5s-6p_{1/2}$ &       &   \\
               & 421.07565(49)  & 421.08(3) \cite{arora11a}    & 421.075(2) \cite{herold12a} \\
$ 5s-6p_{3/2}$ &       &   \\
\end{tabular}
\end{ruledtabular}
\end{table}

Table \ref{tab6} shows the present three tune-out wavelengths of the
$5s_{1/2}$ state of Rb, which are compared with the RMBPT calculations and 
some available experiments. 
In the present calculations of dynamic polarizabilities, 
the matrix elements of $5s-5p_J$ and $5s-6p_J$ transitions are replaced by
the most accurate experimental values \cite{leonard15a,herold12a}. 
There are two cases that the tune-out wavelengths occur.
The first case is that the tune-out wavelength exists between $np_{1/2} $
and $np_{3/2} $ spin-orbit doublet, such as 790.02765 nm lies in the $5s-5p_J$ 
splitting and 421.07565 nm lies in the $5s-6p_J$ splitting. 
The present tune-out wavelength, 790.02765 nm, is shorter 
than the early RMBPT result \cite{arora11a} by 0.007 nm, but agrees with 
the latest RMBPT result 790.0261(7) nm very well. 
There are two experiments \cite{catani09a,lamporesi10a} 
of the longest tune-out wavelength of the $5s$ state available. 
The experiment of Lamporesi $\emph{et al.}$ \cite{lamporesi10a}, 
790.018(2) nm, agrees with 
the RMBPT and the present RCICP theoretical results very well.
The experiment of Catani $\emph{et al.}$ \cite{catani09a}, 
789.85(1)nm, has a big difference
with the available values \cite{leonard15a, arora11a, lamporesi10a} 
and the present RCICP calculation. 
The reason of this difference should be from the
light is not the linearly polarized in this experiment \cite{catani09a}.
The present tune-out wavelength near 421 nm agrees with the RMBPT result \cite{arora11a} and 
experimental result \cite{herold12a} perfectly . 
The second case is that the 
tune-out wavelength occurs when the wavelength is shorter than 
the $5s-np_{3/2}$ transition wavelength and longer than the $5s-(n+1)p_{1/2}$ 
transition wavelength, 
such as 423.02428 nm lies between the $5p_{3/2}$ and $6p_{1/2}$.
This tune-out wavelength also has a good agreement 
with the experimental result \cite{herold12a} and MBPT result \cite{arora11a}.

\section{Results of hyperfine structure}

\subsection{Energies and Reduced matrix elements}

According to first-order perturbation theory, the energy for a hyperfine
state $|LJIF\rangle $ is given \cite{arimondo77a,armstrong71a} by
\begin{eqnarray}
E = E_{NLJ} + W_F,
\end{eqnarray}
where $E_{NLJ}$ is the energy of the unperturbed fine structure state, and $W_F$ is the
hyperfine interaction energy which can be written as
\begin{eqnarray}
W_F = \frac{1}{2}AR+B\frac{\frac{3}{2}R(R+1)-2I(I+1)J(J+1)}{2I(2I-1)2J(2J-1)},
\end{eqnarray}
where $A$ and $B$ are hyperfine
structure constants, and it is usual to give the $A$ and $B$ coefficients in MHz where
$1.0 \ {\rm MHz} = 1.519829903\times 10^{-10}$ a.u..  
\begin{eqnarray}
R=F(F+1)-I(I+1)-J(J+1).
\end{eqnarray}
$F$ is the total angular momentum of the hyperfine state, $I$ is the nuclear spin 
($I=3/2$ for $^{87}$Rb and $I=5/2$ for $^{85}$Rb  ), and 
$J$ is the total angular momentum of the atomic state.

The hyperfine interaction energies of the 
different hyperfine levels of $5s_{1/2}$, $5p_J$ and $6p_J$
states of $^{87,85}$Rb are listed in Table \ref{tab7}.
The hyperfine structure constants A and B are originated from other 
documents \cite{safronova99a,rapol03a,arimondo77a,safronova11aa}.
The energy shifts of the $5s_{1/2}$ state are about one or two order of
magnitude larger than those of the $5p_{J}$, $6p_{J}$ excited states.
Similarly, the hyperfine splittings of the 
$np_{1/2}$ states are obviously larger than the splittings of the 
$np_{3/2}$ states.

\begin{table}
\caption{\label{tab7} The hyperfine interaction energies of the hyperfine states of
$^{85}$Rb and $^{87}$Rb. The notation $a[b]$ means $a \times 10^{b}$. 
Hyperfine structure constants are from other documents.}
\begin{ruledtabular}
\begin {tabular}{lccccr}
     &  J  &    $A$ (MHz)   &  $B$  (MHz)  &  $F$    &  $W_F$ (a.u.)  \\\hline
\multicolumn{5}{c}{$^{87}$Rb, $I = 3/2$ }  \\
$5s$ & 1/2 &  3417.341307 \cite{safronova11aa}   &        & 1   & $-$6.4922[-7]  \\
     &     &    	                        &        & 2   &    3.8953[-7]  \\
$5p$ & 1/2 &  406.2 \cite{safronova11aa}         &        & 1   & $-$7.7169[-8]  \\
     &     &      	                        &        & 2   &    4.6302[-8]  \\
$5p$ & 3/2 &  84.845 \cite{safronova11aa}        &  12.52 \cite{safronova11aa} & 0   & $-$4.5978[-8]  \\
     &     &                                    &        & 1   & $-$3.4986[-8]  \\ 
     &     &                                    &        & 2   & $-$1.1098[-8]  \\
     &     &                                    &        & 3   &    2.9489[-8]  \\
$6p$ & 1/2 &  132.565 \cite{safronova11aa}       &        & 1   & $-$2.5185[-8]  \\
     &     &         	                        &        & 2   &    1.5111[-8]  \\
$6p$ & 3/2 &  27.700 \cite{safronova11aa}        &  3.593 \cite{safronova11aa} & 0   & $-$1.5036[-8]  \\
     &     &                                    &        & 1   & $-$1.1427[-8]  \\ 
     &     &                                    &        & 2   & $-$3.6080[-9]  \\
     &     &                                    &        & 3   &    9.6225[-9]  \\
\multicolumn{5}{c}{$^{85}$Rb, $I = 5/2$ }  \\
$5s$ & 1/2 &   1011.910813  \cite{safronova99a}   &        & 2   & $-$2.6914[-7]  \\
&     &  	            &        & 3   &    1.9224[-7]  \\
$5p$ & 1/2 &   120.7 \cite{safronova99a}   &        & 2   & $-$3.2108[-8]  \\
&     &                &        & 3   &    2.2934[-8]  \\
$5p$ & 3/2 &   25.038 \cite{rapol03a}  & 26.011 \cite{rapol03a} & 1   & $-$1.7218[-8]  \\
&     &                &        & 2   & $-$1.2756[-8]  \\ 
&     &                &        & 3   & $-$3.1143[-9]  \\
&     &                &        & 4   &    1.5248[-8]  \\
$6p$ & 1/2 &   39.11 \cite{safronova99a}       &        & 2   & $-$1.0402[-8]  \\
&     &                &        & 3   &    7.4301[-9]  \\
$6p$ & 3/2 &   8.25  \cite{safronova99a}   & 8.40 \cite{arimondo77a}  & 1   & $-$5.6891[-9]  \\
&     &                &        & 2   & $-$4.2027[-9]  \\ 
&     &                &        & 3   & $-$1.0156[-9]  \\
&     &                &        & 4   &    5.0211[-9]  \\
\end{tabular}
\end{ruledtabular}
\end{table}

The dipole matrix elements 
between the hyperfine states are calculated by using the 
Wigner-Eckart theorem.
The transition matrix elements between the two hyperfine states $ |n_iL_iJ_iIF_i\rangle $ and
$|n_gL_gJ_gIF_g\rangle$ can be written as
\begin{eqnarray}
&&\langle L_gJ_gIF_g \| r^{k}C^{k}(r)\| L_iJ_iIF_i\rangle = (-1)^{I+J_g+F_i+k} \nonumber \\ 
            & \times & \hat{F}_i \hat{F}_g 
  \left\{
\begin{array}{ccc}
I & J_i & F_i \\
k & F_g & J_g \\
\end{array}
\right \}
\langle L_gJ_g \| r^{k}C^{k}(r) \| L_iJ_i \rangle ,
\end{eqnarray}
where $k=1$ for a dipole transition and $\hat{F} = \sqrt{2F+1}$.

The absorption oscillator strength $f^{(k)}_{gi}$ for
a transition from hyperfine state $g \to i$
is defined as
\begin{equation}
f^{(k)}_{gi} = \frac{ 2|\langle L_iJ_iIF_i \| r^{k}C^{k}(r) \| L_gJ_gIF_g\rangle 
|^2 \varepsilon_{gi} } {(2k+1)(2F_g+1)} \ .
\end{equation}

In the present calculations, in order to consider  
energy dependent correction of the matrix elements, 
the matrix elements are treated 
as parametric functions of their
binding energies \cite{jiang13b}. The functional form is
\begin{eqnarray}
A_{ij}(E_i,E_j) &\approx&  A_{ij}(E_{0,i},E_{0,j}) + \frac{\partial A_{ij}}{\partial E_i}(E_i-E_{0,i})
         \nonumber \\ 
	 &+& \frac{\partial A_{ij}}{\partial E_j}(E_j-E_{0,j}) \ , 
\end{eqnarray}
where $E_{0,i}$ and $E_{0,j}$ are the binding energies without any hyperfine splitting. 
The partial derivatives are evaluated 
by redoing the calculations with the slightly different polarization 
potentials and leading to the change in the reduced matrix elements. 
The partial derivatives of matrix elements are listed 
in Table \ref{tab8}. 

\begin{table}
\caption{\label{tab8} The partial derivatives for the matrix elements of $5s-5p_J$ and
$5s-6p_J$ transitions with respect to the initial and final state binding
energies.  }
\begin{ruledtabular}
\begin {tabular}{lcc}
Transition  & $\frac{\partial A}{\partial E_{5s}}$  & $\frac{\partial A}{\partial E_{j}} $  \\  \hline
  $5s_{1/2}$-$5p_{1/2}$   & 31.070953     & $-$1.800089       \\
  $5s_{1/2}$-$5p_{3/2}$   & 44.794208     & $-$4.888183       \\
  $5s_{1/2}$-$6p_{1/2}$   & $-$17.415952  & 136.505937          \\
  $5s_{1/2}$-$6p_{3/2}$   & $-$23.446857  & 208.756559          \\
\end{tabular}
\end{ruledtabular}
\end{table}

\subsection{Dipole Polarizabilities of the hyperfine ground states}

The dynamic dipole polarizabilities are computed with the 
usual oscillator strength sum-rules in Eq.(8), 
where the sum over $n$ includes all allowable hyperfine structure transitions. 
In the calculations of polarizabilities for the hyperfine states, 
the resonance transition energies of hyperfine levels 
of the $5s, 5p_{J}, 6p_{J}$ states are replaced by the 
experimental results \cite{steck10a}. The  
uncertainties of these resonance transition 
energies reach to $ 3.8 \times 10^{-8} $ eV.

The dipole polarizability also has a tensor component 
for states with $ F > 1/2 $.  It can be written as
\begin{eqnarray}
&&\alpha_{\rm T}^{(1)}(\omega)=6\left( \frac{5F_g(2F_g-1)(2F_g+1)}{6(F_g+1)(2F_g+3)}
 \right)^{1/2} \nonumber \\
                           & \times & \sum_{i}  (-1)^{F_g+F_i}
      \left\{
\begin{array}{ccc}
F_g & 1   & F_i \\
  1 & F_g & 2  \\
\end{array} 
\right\}
\frac{ f^{(1)}_{gi} } {\varepsilon_{gi}^2 - \omega^2}.
\end{eqnarray}
The dipole polarizabilities of the hyperfine levels 
can be calculated by the following equation \cite{mitroy10a}, 
\begin{equation}
\alpha^{(1)}_{M_g}(\omega) = \alpha^{(1)}(\omega) + \alpha_{T}^{(1)}(\omega) \frac{3M_g^2-F_g(F_g+1)}{F_g(2F_g-1)}. 
\label{polar3}
\end{equation}
\begin{table}
\caption{\label{tab9} The scalar $\alpha^{(1)}$ and tensor $\alpha_{T}^{(1)}$ dipole
polarizabilities of the hyperfine ground states of $^{87,85}$Rb. The notation $a[b]$ 
means $a \times 10^{b}$. 
}
\begin{ruledtabular}
\begin {tabular}{llccc}
          & State      & $F$ & $\alpha^{(1)}$ (a.u.) &    $ \alpha_{T}^{(1)}$ (a.u.) \\
\hline
$^{87}$Rb & $5s_{1/2}$ &   1   &   318.699491    &    1.5883[$-$5]     \\
          & $5s_{1/2}$ &   2   &   318.709441    & $-$8.8203[$-$5]   \\
\hline
$^{85}$Rb & $5s_{1/2}$ &   2   &   318.702958    &    2.0494[$-$5]     \\
          & $5s_{1/2}$ &   3   &   318.707444    & $-$4.0621[$-$5]    \\
\end{tabular}
\end{ruledtabular}
\end{table}

\begin{table}
\caption{\label{tab10} The difference of scalar and tensor dipole polarizabilities
of the hyperfine ground states of $^{87,85}$Rb.
The notation $a[b]$ means $a \times 10^{b}$. 
}
\begin{ruledtabular}
\begin {tabular}{lc}
Method           &  $\Delta \alpha^{(1)}$ (a.u.)     \\ \hline
&  \multicolumn{1}{c}{$^{87}$Rb: $\alpha^{(1)}(F=2)-\alpha^{(1)}(F=1)$} \\\hline
Present RCICP                        & 0.995[$-$2]   \\
RCI + MBPT \cite{angstmann06a}       & 0.997(8)[$-$2] \\
RLCCSDT \cite{safronova10a}          & 0.997(3)[$-$2] \\
Perturbation theory \cite{lee75a}    & 0.972[$-$2]    \\
Expt. \cite{mowat72a}                & 0.99(24)[$-$2]  \\
Expt. \cite{dallal15a}               & 0.9967(32)[$-$2] \\\hline
&  \multicolumn{1}{c}{$^{85}$Rb:  $\alpha^{(1)}(F=3)-\alpha^{(1)}[(F=2)$} \\\hline
Present RCICP                        & 4.486[$-$3]  \\
Perturbation theory \cite{lee75a}    & 4.311[$-$3]   \\
Expet. \cite{mowat72a}               & 4.389(96)[$-$3] \\\hline
&  \multicolumn{1}{c}{$^{87}$Rb:  $\alpha^{(1)}_T(F=2)-\alpha^{(1)}_T(F=1)$} \\\hline
Present RCICP                        & $-$1.0409[$-$4]   \\
Expt. \cite{dallal15a}               & $-$0.8841(1045)[$-$4] \\\hline
&  \multicolumn{1}{c}{$^{85}$Rb:  $\alpha^{(1)}_T(F=3)-\alpha^{(1)}_T(F=2)$} \\\hline
Present RCICP                        & $-$6.1115[$-$5]   \\
\end{tabular}
\end{ruledtabular}
\end{table}

Table \ref{tab9} gives the static scalar and tensor dipole polarizabilities 
of the hyperfine ground states of $^{87, 85}$Rb. There are no other 
theoretical or experimental results that can be directly 
compared with these values in Table \ref{tab9}. 
However, the hyperfine stark shift, which is
the difference of scalar polarizabilities between the hyperfine 
states with the same $(L,J)$ but different $F$ quantum numbers, can be 
compared with other theoretical and experimental results. 
Table \ref{tab10} gives the differences of scalar and tensor polarizabilities
of the hyperfine ground states of $^{87, 85}$Rb in a.u..  
There are some documents of the hyperfine Stark shifts of $^{87, 85}$Rb
 \cite{lee75a,mowat72a,angstmann06a,safronova10a,dallal15a} that are often 
reported as the Stark shift coefficients $k$, with units of (Hz/(V/m)$^2$. 
This is converted into a.u. by multiplying 0.4018778$\times 10^{8}$ \cite{mitroy10a}. 
The present hyperfine Stark shift of $^{87}$Rb is slightly smaller than 
the relativistic configuration interaction plus many-body 
perturbation(RCI + MBPT) \cite{angstmann06a}, 
the relativistic linearized coupled cluster single-double 
with partial triple contributions(RLCCSDT) \cite{safronova10a}, 
and larger than the perturbation theory \cite{lee75a}.
This value is also between the experimental value \cite{mowat72a} by Mowat $\emph{et al.}$ 
and the experimental value \cite{dallal15a} by Dallal $\emph{et al.}$.
The present hyperfine Stark shift of $^{85}$Rb is larger than 
Perturbation theory \cite{lee75a} and experimental value \cite{mowat72a}.

The tensor polarizabilities of the hyperfine states do not 
exceed $ 10^{-4}$ a.u. in magnitude. 
The tensor polarizability of the $ F=1 $ ground state of $^{87}$Rb is positive and 
that of the $ F=2 $ ground state of $^{87}$Rb is negative.
The difference of present tensor polarizabilities of 
the $ F=2 $ and $ F=1 $ ground states of $^{87}$Rb 
is $-1.0409 \times 10^{-4}$ a.u., which is  
more negative than the experimental value $-0.8841 \times 10^{-4}$ a.u. \cite{dallal15a}. 
The difference between experiment and the present calculation 
is $1.568 \times 10^{-5}$ a.u., which 
is larger than the experimental error bar $1.045 \times 10^{-5}$ a.u..  
The tensor polarizability of the $ F=2 $ ground state of $^{85}$Rb is positive and 
that of the $ F=3 $ ground state of $^{85}$Rb is negative. 
The difference of tensor polarizabilities of the 
$ F=3 $ and $ F=2 $ ground states of $^{85}$Rb  
is $-6.1115 \times 10^{-5}$ a.u.. 
There are no any other comparable theoretical and experimental data available at present. 

The energy-dependent corrections of the dipole matrix elements play an important
role in the calculation of dynamic polarizabilities. 
Omitting the matrix element correction results in the
hyperfine Stark shifts about half these values, namely , 5.454 $\times 10^{-3}$ a.u.
of $^{87}$Rb and 2.423 $\times 10^{-3}$ a.u. of $^{85}$Rb respectively.

\subsection{Tune-out wavelengths of the hyperfine ground states}

\subsubsection{ $^{87}$Rb }

\begin{table}
\caption{\label{tab11} Tune-out wavelengths $\lambda_{zero}$ (in nm) of the
$5s_{1/2},F=1$ and $5s_{1/2},F=2$ states of $^{87}$Rb.
$ \Delta \lambda $ (in nm) is the shifts of the primary tune-out wavelengths 
compared to the tune-out wavelengths of $5s$ state.
Tune-out wavelengths are given to six digits after the decimal point.}
\begin{ruledtabular}
\begin {tabular}{cccc}
\multicolumn{2}{c}{F=1}  &  \multicolumn{2}{c}{F=2}     \\
\cline{1-2} \cline{3-4}
$\lambda_{zero}$ & $10^{-3} \Delta \lambda $ & $\lambda_{zero}$ & $10^{-3} \Delta \lambda $ \\\hline
794.970633  & -  &  794.984469  & - \\
790.018187  & $-$9.46  &  790.032602  & +4.95 \\
780.233113  & -  &  780.246852  & - \\
780.232827  & -  &  780.246413  & - \\
423.021740  & $-$2.42  &  423.025808  & +1.46 \\
421.670240  & -  &  421.674241  & - \\
421.073131  & $-$2.51  &  421.077158  & +1.51 \\
420.296547  & -  &  420.300560  & - \\
420.296519  & -  &  420.300519  & - \\
\end{tabular}
\end{ruledtabular}
\end{table}

Hyperfine splittings lead to two new features in the tune-out wavelengths. 
One feature is that the splitting of the $5s_{1/2}$ state
has resulted in two duplicate sets of tune-out wavelengths, 
that is for the $F = 1$ and $F = 2$ hyperfine ground states. 
Another feature is that the hyperfine splittings of the $5p_{J,F}$ state 
have also resulted in
the creation of additional tune-out wavelengths that
arise from two adjacent hyperfine states each other.
The hyperfine splitting of the $5p_{1/2}$ state has resulted in one additional tune-out 
wavelength, located between the $5p_{1/2}, F=1$ and $5p_{1/2},F=2$
states. The hyperfine structure with regard to $5p_{3/2}$ state brings
two additional tune-out wavelengths, located between the three $5p_{3/2}, F=1,2,3$
levels with allowed dipole transitions to the $5s_{1/2},F=2$
hyperfine state, or between the three $5p_{3/2}, F=0,1,2$
levels with allowed dipole transitions to the $5s_{1/2},F=1$ hyperfine state. 
There are several tune-out wavelengths that are defined as the
primary tune-out wavelengths, which are the closest
to the tune-out wavelengths calculated without the hyperfine splittings.

Table \ref{tab11} gives the tune-out wavelengths 
of the two hyperfine ground states of the $5s_{1/2}$ state of $^{87}$Rb.
These wavelengths are given to six digits after the decimal point 
to ensure that all the differences 
of the tune-out wavelengths are at least two digits.
The longest tune-out wavelengths near 794 nm occur in the hyperfine splitting 
of the $5p_{1/2}$ state. These tune-out wavelengths would be hard to detect due to
the very small energy splittings of the hyperfine states. 
The second tune-out wavelengths near 790 nm are
the first primary tune-out wavelengths, which lie between the excitation
thresholds of the $5p_{1/2}$ and $5p_{3/2}$ states.
The first primary tune-out wavelengths 
of the $5s_{1/2}, F = 1, 2$ states are 790.018187 nm and 790.032602 nm respectively. 
The present calculation, 790.032602 nm of the $5s_{1/2}, F = 2$ state, is larger than 
the latest experiment 790.032388(32) nm \cite{leonard15a} 
and the difference is 0.000214 nm. This difference is still about seven times 
larger than the experimental error bar. 
The tune-out wavelengths near 423 nm are other primary tune-out wavelengths, 
which lie between the excitation
thresholds of the $5p_{3/2}$ and $6p_{1/2}$ states.
Similarly, the tune-out wavelengths near 421.07 nm
are also primary tune-out wavelengths, 
which lie between the excitation
thresholds of the $6p_{1/2}$ and $6p_{3/2}$ states. 
These primary tune-out wavelengths of the $5s_{1/2}, F = 1$ state
are shorter than the corresponding tune-out wavelengths of $5s$ state of Rb,
and those of the $5s_{1/2}, F = 2$ state 
are longer than the close tune-out wavelengths of $5s$ state of Rb.  
The tune-out wavelengths near 780 nm, 421.67 nm, and 420.3 nm occur 
in the hyperfine splittings of the $5p_{3/2}$, $6p_{1/2}$, and $6p_{3/2}$ states respectively,
which are also very hard to detect. 

\begin{table*}
\caption{\label{tab12} Tune-out wavelengths $\lambda_{zero}$ (in nm) of the different magnetic sublevels
of $5s_{1/2},F=1$ and $5s_{1/2},F=2$ states of $^{87}$Rb.
Tune-out wavelengths are given to seven digits after the decimal point.}
\begin{ruledtabular}
\begin {tabular}{ccccc}
\multicolumn{2}{c}{F=1}  &  \multicolumn{3}{c}{F=2}     \\
\cline{1-2} \cline{3-5}
$ M_{F}=-1,1$  & $ M_{F}=0 $   & $ M_{F}=-2,2$   & $ M_{F}=-1,1 $   & $ M_{F}=0 $  \\
\hline
794.9705853    & 794.9707284   & 794.9846000     & 794.9844029      & 794.9843373  \\
790.0182169    & 790.0181259   & 790.0325203     & 790.0326434      & 790.0326845  \\ 
780.2331259    & 780.2330860   & 780.2468572     & 780.2468488      & 780.2468458  \\
780.2328185    & 780.2328473   & 780.2463937     & 780.2464233      & 780.2464334  \\
423.0217422    & 423.0217345   & 423.0258015     & 423.0258118      & 423.0258153  \\
421.6702358    & 421.6702489   & 421.6742534     & 421.6742354      & 421.6742294  \\ 
421.0731328    & 421.0731283   & 421.0771545     & 421.0771603      & 421.0771623  \\
420.2965477    & 420.2965439   & 420.3005609     & 420.3005601      & 420.3005598  \\
420.2965184    & 420.2965212   & 420.3005169     & 420.3005197      & 420.3005206  \\
\end{tabular}
\end{ruledtabular}
\end{table*}

The tune-out wavelengths also depend on the magnetic
sublevels if tensor polarizabilities are considered. The tune-out wavelengths associated
with the different magnetic sublevels of the $5s_{1/2},F$ states of $^{87}$Rb are listed
in Table \ref{tab12}. Comparing with the tune-out wavelengths 
for the different magnetic sublevels of the same hyperfine ground state,  
the shifts of tune-out wavelengths due to tensor polarizabilities are less than 
$10^{-4}$ nm. Here we focus on that the difference of the first 
primary tune-out wavelengths for the $M_F= \pm{1}$ and $M_F=0$ of 
the $5s_{1/2},F=1$ state is $9.1 \times 10^{-5}$ nm.
The first primary tune-out wavelength of the  
$M_F = 0$ sublevel of $5s_{1/2},F=1$ state is 790.0181259 nm. 
It is a little shorter than the 
very recent experiment 790.01858(23) nm \cite{schmidt16a},
and the difference is about 0.00045 nm which
is nearly two times larger than the experimental error bar.
The first primary tune-out wavelengths for the  $M_F = 0$, $M_F = \pm{1}$
and $M_F = \pm{2}$ sublevels of the $5s_{1/2}, F=2$ state are 790.0326845 nm, 790.0326434 nm
and 790.0325203 nm respectively. The differences in these tune-out wavelengths for any of
the different magnetic sublevels do not exceed $1.7 \times 10^{-4}$ nm.

\subsubsection{$^{85}$Rb}

\begin{table}
\caption{\label{tab13} Tune-out wavelengths $\lambda_{zero}$ (in nm) of the
$5s_{1/2}, F=2$ and $5s_{1/2}, F=3$ states of $^{85}$Rb.
$ \Delta \lambda $ (in nm) is the shifts of the primary tune-out wavelengths 
compared to the tune-out wavelengths of $5s$ state.
Tune-out wavelengths are given to six digits after the decimal point.}
\begin{ruledtabular}
\begin {tabular}{cccc}
\multicolumn{2}{c}{F=2}  &  \multicolumn{2}{c}{F=3}     \\
\cline{1-2} \cline{3-4}
$\lambda_{zero}$      & $10^{-3} \Delta \lambda $ &    $\lambda_{zero}$ & $10^{-3} \Delta \lambda $   \\\hline
794.975393  & -  &  794.981538  & - \\
790.023515  & $-$0.41  &  790.029918 & +2.27 \\
780.237979  & -  &  780.244089  & - \\
780.237860  & -  &  780.243899  & - \\
423.023277  & $-$1.00  &  423.025001 & +0.72 \\
421.671676  & -  &  421.673454  & - \\
421.074607  & $-$1.04  &  421.076392 & +0.74  \\
420.297985  & -  &  420.299769  & - \\
420.297974  & -  &  420.299750  & - \\
\end{tabular}
\end{ruledtabular}
\end{table}

Table \ref{tab13} gives the tune-out wavelengths of the
$5s_{1/2},F=2$ and $5s_{1/2},F=3$ states of $^{85}$Rb. 
Table \ref{tab14} gives the tune-out wavelengths for the different magnetic sublevels. 
All analysis and properties of $^{85}$Rb should be interpreted with the contents
of the previous section in mind.  
The differences between the tune-out wavelengths  
of the hyperfine states of $^{85}$Rb are smaller than
those of $^{87}$Rb. 
It is understandable since $^{85}$Rb has the smaller hyperfine
structure constants than $^{87}$Rb.
Similarly, the differences between the tune-out wavelengths  
of the hyperfine magnetic sublevels 
of $^{85}$Rb are also smaller than
those of $^{87}$Rb.

\begin{table*}
\caption{\label{tab14} Tune-out wavelengths $\lambda_{zero}$ (in nm) of the different magnetic sublevels of
$5s_{1/2},F=2$ and $5s_{1/2},F=3$ states of $^{85}$Rb.
Tune-out wavelengths are given to seven digits after the decimal point.}
\begin{ruledtabular}
\begin {tabular}{ccccccc}
\multicolumn{3}{c}{F=2}    &  \multicolumn{4}{c}{F=3}     \\
\cline{1-2}  \cline{3-7}
$ M_{F}=-2,2$  & $ M_{F}=-1,1 $  & $ M_{F}=0  $  & $ M_{F}=-3,3$   & $ M_{F}=-2,2 $  & $ M_{F}=-1,1 $   & $ M_{F}=0 $ \\
\hline
794.9753668    & 794.9754056     & 794.9754185   & 794.9815835     & 794.9815382     & 794.9815110      & 794.9815020 \\
790.0235316    & 790.0236506     & 790.0234980   & 790.0298900     & 790.0299180     & 790.0299347      & 790.0299403 \\
780.2379826    & 780.2379764     & 780.2379743   & 780.2440913     & 780.2440888     & 780.2440870      & 780.2440865 \\
780.2378566    & 780.2378612     & 780.2378629   & 780.2438904     & 780.2438990     & 780.2439042      & 780.2439060 \\
423.0232785    & 423.0232769     & 423.0232764   & 423.0250000     & 423.0250014     & 423.0250024      & 423.0250028 \\
421.6716741    & 421.6716777     & 421.6716789   & 421.6734579     & 421.6734538     & 421.6734513      & 421.6734505 \\
421.0746085    & 421.0746070     & 421.0746065   & 421.0763902     & 421.0763917     & 421.0763927      & 421.0763930 \\
420.2979853    & 420.2979847     & 420.2979845   & 420.2997688     & 420.2997685     & 420.2997683      & 420.2997683 \\
420.2979732    & 420.2979737     & 420.2979738   & 420.2997495     & 420.2997503     & 420.2997508      & 420.2997510 \\
\end{tabular}
\end{ruledtabular}
\end{table*}

\subsection{Some comments on accuracy}

The uncertainties of the dipole reduced matrix elements
of the $5s-5p_J$ transitions are mainly caused by
the correlation effects of frozen-core model.
These uncertainties are smaller than 0.5$\%$,   
thus we set 0.5$\%$ as the uncertainties of the
dipole reduced matrix elements. 
The uncertainties of dipole reduced matrix elements for 
the transitions of more highly-excited states are derived from 
the first-order parametric functions of their energies.
By considering the uncertainties of dipole reduced matrix elements, 
the uncertainties of three tune-out wavelengths of the ground state of Rb 
are obtained.

Comparing with the present calculations and available experimental results \cite{leonard15a,schmidt16a}, 
the absolute precision of tune-out wavelengths should be about 0.0005 nm. 
The method used to determine the tune-out wavelengths of hyperfine states was 
unorthodox, being essentially a second order calculation using energy 
and matrix element shifts applied prior to the evaluation of the oscillator 
strength sum-rules. There are three main factors 
that influence the accuracy of the present tune-out wavelengths. 
Table \ref{tab15} shows the estimated errors of the first primary tune-out wavelengths 
of the $5s_{1/2}, F=1, 2$ states of $^{87}$Rb and $5s_{1/2}, F=2, 3$ states of $^{85}$Rb.

\begin{table}
\caption{\label{tab15} The estimated errors (in fm) of the first primary tune-out wavelengths 
of the $5s_{1/2}, F=1, 2$ states of $^{87}$Rb and $5s_{1/2}, F=2, 3$ states of $^{85}$Rb.
$\delta\lambda_1$ is the errors that are caused by the 0.05 $\%$ uncertainties of 
$5s-5p_{1/2}$ and $5s-5p_{3/2}$ matrix elements. 
$\delta\lambda_2$ is the errors that are caused by the 0.00003 
uncertainty of ratio of $5s-5p_J$ line strengths.
$\delta\lambda_3$ is the errors that are caused by 5$\%$ uncertainties of the 
matrix elements from the high-excited, continumm, and core-excited states.
$\sum\delta\lambda_i$ is the sum of the $\delta\lambda_1$, $\delta\lambda_2$, and $\delta\lambda_3$. 
}
\begin{ruledtabular}
\begin {tabular}{ccccccc}
            & F &    $\lambda_{zero}$(nm) & $\delta\lambda_1$ &  $\delta\lambda_2$ & $\delta\lambda_3$ &  $ \sum\delta\lambda_i $\\\hline
 $^{87}$Rb  & 1 &   790.0181865   &    5    &   13   &  175   & 193 \\
            & 2 &   790.0326024   &    5    &   13   &  175   & 193 \\
 $^{85}$Rb  & 2 &   790.0235148   &    5    &   13   &  200   & 218 \\
            & 3 &   790.0299179   &    5    &   13   &  200   & 218 \\    
\end{tabular}
\end{ruledtabular}
\end{table}

The first factor is the uncertainties of $5s-5p_J$ matrix elements.
An uncertainty analysis has been done for the tune-out wavelengths.
Firstly, the matrix elements of $5s-5p_{1/2}$ and $5s-5p_{3/2}$ 
transitions are changed by 0.05\% according to the errors 
between the present RCICP calculations and Ref. \cite{leonard15a}.
The matrix elements are adjusted accordingly 
and tune-out wavelengths are recomputed.
In this case, the ratio of line strengths $5s-5p_J$ is not changed. 
The shifts of the first primary tune-out wavelengths near 790.0 nm of $ ^{87,85}$Rb are about 5 fm. 
It is too small to explain the differences 
of the present calculations and experimental results \cite{leonard15a,schmidt16a}.
The shifts of the other primary tune-out wavelengths 
near 423.0 nm which lie in $5s-5p_{3/2}$
and $5s-6p_{1/2}$ transitions of $ ^{87,85}$Rb are 2034 fm. 
The shifts of the primary tune-out wavelengths near 421.0 nm 
which lie in $5s-6p_{1/2}$
and $5s-6p_{3/2}$ transitions of $ ^{87,85}$Rb are about 257 fm.
The shifts of the tune-out wavelengths which lie in the $np_J$ hyperfine splittings 
are smaller than $10^{-11}$ nm. 
Then, the ratio of line strengths $5s-5p_J$ is changed by 0.00003. 
The shifts of the first primary tune-out wavelengths of $ ^{87,85}$Rb 
near 790.0 nm are 13 fm. 
These shifts are still much smaller than the differences  
between the present calculations and latest experiments \cite{leonard15a,schmidt16a}.
The shifts of the other primary tune-out wavelengths of $^{87,85}$Rb
near 423.0 nm are about 6 fm. 
The shifts of the primary tune-out wavelengths of $^{87,85}$Rb 
near 421.0 nm are about 0.7 fm.

We also have checked the sensitivity of the tune-out wavelengths 
to the small changes in the energy-adjusted matrix elements.
The tune-out wavelengths are recalculated 
without the modifications of matrix elements due to the energy-adjustment. 
The tune-out wavelengths are insensitive to these small changes,
these are totally different with 
the hyperfine stark shifts which are critically reliant on 
the use of energy-adjusted matrix elements. 
For example, the energy-adjusted reduced matrix elements
make the 0.5 fm shifts to the first primary tune-out wavelengths of the
$F = 1,2$ ground states of $^{87}$Rb.
The shifts of first primary tune-out wavelengths
of the $F=2,3$ ground states of $^{85}$Rb are about 0.2 fm.  
These shifts are two or three orders smaller than latest 
experimental error bars \cite{leonard15a,schmidt16a}.

The second factor is the uncertainties in contributions to polarizabilities
from the highly-excited, continuum, and core-excited states.
The contribution from excited states above $5p$ state is 
11.14 a.u. in the present RCICP calculations of tune-out wavelengths, 
which is $4.1\%$ difference with the 
value given by Leonard $\emph{et al.}$ \cite{leonard15a}.
So we changed the matrix elements of highly-excited states 
by 5$\%$ and the tune-out wavelengths are recomputed. 
The first primary tune-out wavelengths 
will shift 175 fm of $^{87}$Rb and 200 fm of $^{85}$Rb.
These shifts are close to the difference between the present 
calculation and the latest experiment \cite{leonard15a}. 

The third factor is the uncertainties of transition energies of hyperfine states. In the 
present calculations, the experimental resonance transition energies \cite{steck10a} 
for the hyperfine transitions are used, 
in which the uncertainties are smaller than $10^{-7}$ nm. 
The effect of hyperfine structure for the level higher than $6p$ is negligible. 
So this factor can be ignored in the present analysis. 

\section{Conclusions}  

The static and dynamic polarizabilities of the ground state of Rb 
have been calculated by using the RCICP method. 
Combining the most exact $5s-5p_J$ matrix elements \cite{leonard15a},
three longest tune-out wavelengths of the $ 5s_{1/2} $ state are determined. 
After considering the hyperfine splittings of energy levels,
the static and dynamic polarizabilities, 
and the tune-out wavelengths of the hyperfine ground states of $^{87,85}$Rb 
are further determined. The present hyperfine stark shifts are in good agreement with 
the available theoretical and experimental results. 
Considering the contributions of tensor polarizabilities, 
the tune-out wavelengths for the different magnetic 
sublevels $ M_{F}$ of the hyperfine states $F$ are obtained. 
It is found that the differences of the tune-out wavelengths 
for the different magnetic sublevels do not exceed $10^{-4}$ nm.

The first primary tune-out wavelengths of the $ 5s_{1/2}, F=1, 2 $ 
states of $ ^{87}$Rb are 790.018187(193) nm and 790.032602(193) nm severally.
The first primary tune-out wavelengths of the $ 5s_{1/2}, F=2, 3 $ 
states of $ ^{85}$Rb are 790.023515(218) nm and 790.029918(218) nm respectively.
The present results are compared with the recent experiments \cite{leonard15a,schmidt16a}. 
The differences between the present calculations 
and the recent experiments are still larger than the experimental error bars \cite{leonard15a,schmidt16a}.
But the present RCICP first primary tune-out wavelength of  
$ 5s_{1/2}, F= 2 $ state of $ ^{87}$Rb
is longer than that observed in the recent experiment \cite{leonard15a}. 
Meanwhile, the present RCICP first primary tune-out wavelength of 
$ 5s_{1/2}, F= 1, M_F = 0 $ state of $ ^{87}$Rb
is shorter than that observed in the latest experiment \cite{schmidt16a}. 
It seems that the main uncertainty of the polarizabilities from the highly-excited, 
continuum and core-excited states can not explain this difference completely, 
because the uncertainty of the remaining polarizabilities can only lead to consisitently
longer or consisitently shorter, but not to some longer and some shorter than 
the first primary tune-out wavelengths in recent experiments\cite{leonard15a,schmidt16a}.
Hence, a further study will be essential.

\begin{acknowledgments}
The work of JJ was supported by
National Natural Science Foundation of China (NSFC) (Grants No.11147018, 11564036).
The work of LYX was supported by NSFC (Grants No. U1331122). 
The work of DHZ was supported by NSFC (Grants No. 11464042,U1330117).
The work of CZD was supported by NSFC (Grants No. 11274254, U1332206). 
\end{acknowledgments}


\begin{thebibliography}{50}
\expandafter\ifx\csname natexlab\endcsname\relax\def\natexlab#1{#1}\fi
\expandafter\ifx\csname bibnamefont\endcsname\relax
  \def\bibnamefont#1{#1}\fi
\expandafter\ifx\csname bibfnamefont\endcsname\relax
  \def\bibfnamefont#1{#1}\fi
\expandafter\ifx\csname citenamefont\endcsname\relax
  \def\citenamefont#1{#1}\fi
\expandafter\ifx\csname url\endcsname\relax
  \def\url#1{\texttt{#1}}\fi
\expandafter\ifx\csname urlprefix\endcsname\relax\def\urlprefix{URL }\fi
\providecommand{\bibinfo}[2]{#2}
\providecommand{\eprint}[2][]{\url{#2}}

\bibitem[{\citenamefont{Safronova
  et~al.}(2013{\natexlab{a}})\citenamefont{Safronova, Safronova, and
  Porsev}}]{safronova13aa}
\bibinfo{author}{\bibfnamefont{M.~S.} \bibnamefont{Safronova}},
  \bibinfo{author}{\bibfnamefont{U.~I.} \bibnamefont{Safronova}},
  \bibnamefont{and} \bibinfo{author}{\bibfnamefont{S.~G.}
  \bibnamefont{Porsev}}, \bibinfo{journal}{Phys. Rev. A}
  \textbf{\bibinfo{volume}{87}}, \bibinfo{pages}{032513}
  (\bibinfo{year}{2013}{\natexlab{a}}),
  \urlprefix\url{http://link.aps.org/doi/10.1103/PhysRevA.87.032513}.

\bibitem[{\citenamefont{Safronova
  et~al.}(2013{\natexlab{b}})\citenamefont{Safronova, Porsev, Safronova,
  Kozlov, and Clark}}]{safronova13b}
\bibinfo{author}{\bibfnamefont{M.~S.} \bibnamefont{Safronova}},
  \bibinfo{author}{\bibfnamefont{S.~G.} \bibnamefont{Porsev}},
  \bibinfo{author}{\bibfnamefont{U.~I.} \bibnamefont{Safronova}},
  \bibinfo{author}{\bibfnamefont{M.~G.} \bibnamefont{Kozlov}},
  \bibnamefont{and} \bibinfo{author}{\bibfnamefont{C.~W.} \bibnamefont{Clark}},
  \bibinfo{journal}{Phys. Rev. A} \textbf{\bibinfo{volume}{87}},
  \bibinfo{pages}{012509} (\bibinfo{year}{2013}{\natexlab{b}}).

\bibitem[{\citenamefont{Porsev and Derevianko}(2006)}]{porsev06aa}
\bibinfo{author}{\bibfnamefont{S.~G.} \bibnamefont{Porsev}} \bibnamefont{and}
  \bibinfo{author}{\bibfnamefont{A.}~\bibnamefont{Derevianko}},
  \bibinfo{journal}{Phys. Rev. A} \textbf{\bibinfo{volume}{74}},
  \bibinfo{pages}{020502} (\bibinfo{year}{2006}),
  \urlprefix\url{http://link.aps.org/doi/10.1103/PhysRevA.74.020502}.

\bibitem[{\citenamefont{Middelmann et~al.}(2012)\citenamefont{Middelmann,
  Falke, Lisdat, and Sterr}}]{middelmann12a}
\bibinfo{author}{\bibfnamefont{T.}~\bibnamefont{Middelmann}},
  \bibinfo{author}{\bibfnamefont{S.}~\bibnamefont{Falke}},
  \bibinfo{author}{\bibfnamefont{C.}~\bibnamefont{Lisdat}}, \bibnamefont{and}
  \bibinfo{author}{\bibfnamefont{U.}~\bibnamefont{Sterr}},
  \bibinfo{journal}{Phys. Rev. Lett.} \textbf{\bibinfo{volume}{109}},
  \bibinfo{pages}{263004} (\bibinfo{year}{2012}),
  \urlprefix\url{http://link.aps.org/doi/10.1103/PhysRevLett.109.263004}.

\bibitem[{\citenamefont{Cheng et~al.}(2013)\citenamefont{Cheng, Jiang, and
  Mitroy}}]{cheng13aa}
\bibinfo{author}{\bibfnamefont{Y.}~\bibnamefont{Cheng}},
  \bibinfo{author}{\bibfnamefont{J.}~\bibnamefont{Jiang}}, \bibnamefont{and}
  \bibinfo{author}{\bibfnamefont{J.}~\bibnamefont{Mitroy}},
  \bibinfo{journal}{Phys. Rev. A} \textbf{\bibinfo{volume}{88}},
  \bibinfo{pages}{022511} (\bibinfo{year}{2013}),
  \urlprefix\url{http://link.aps.org/doi/10.1103/PhysRevA.88.022511}.

\bibitem[{\citenamefont{{LeBlanc} and {Thywissen}}(2007)}]{leblanc07a}
\bibinfo{author}{\bibfnamefont{L.~J.} \bibnamefont{{LeBlanc}}}
  \bibnamefont{and} \bibinfo{author}{\bibfnamefont{J.~H.}
  \bibnamefont{{Thywissen}}}, \bibinfo{journal}{\pra}
  \textbf{\bibinfo{volume}{75}}, \bibinfo{eid}{053612} (\bibinfo{year}{2007}).

\bibitem[{\citenamefont{{Holmgren} et~al.}(2012)\citenamefont{{Holmgren},
  {Trubko}, {Hromada}, and {Cronin}}}]{holmgren12a}
\bibinfo{author}{\bibfnamefont{W.~F.} \bibnamefont{{Holmgren}}},
  \bibinfo{author}{\bibfnamefont{R.}~\bibnamefont{{Trubko}}},
  \bibinfo{author}{\bibfnamefont{I.}~\bibnamefont{{Hromada}}},
  \bibnamefont{and} \bibinfo{author}{\bibfnamefont{A.~D.}
  \bibnamefont{{Cronin}}}, \bibinfo{journal}{Phys.~Rev.~Lett.}
  \textbf{\bibinfo{volume}{109}}, \bibinfo{eid}{243004} (\bibinfo{year}{2012}).

\bibitem[{\citenamefont{Schmidt et~al.}(2016)\citenamefont{Schmidt, Mayer,
  Hohmann, Lausch, Kindermann, and Widera}}]{schmidt16a}
\bibinfo{author}{\bibfnamefont{F.}~\bibnamefont{Schmidt}},
  \bibinfo{author}{\bibfnamefont{D.}~\bibnamefont{Mayer}},
  \bibinfo{author}{\bibfnamefont{M.}~\bibnamefont{Hohmann}},
  \bibinfo{author}{\bibfnamefont{T.}~\bibnamefont{Lausch}},
  \bibinfo{author}{\bibfnamefont{F.}~\bibnamefont{Kindermann}},
  \bibnamefont{and} \bibinfo{author}{\bibfnamefont{A.}~\bibnamefont{Widera}},
  \bibinfo{journal}{Phys. Rev. A} \textbf{\bibinfo{volume}{93}},
  \bibinfo{pages}{022507} (\bibinfo{year}{2016}),
  \urlprefix\url{http://link.aps.org/doi/10.1103/PhysRevA.93.022507}.

\bibitem[{\citenamefont{Leonard et~al.}(2015)\citenamefont{Leonard, Fallon,
  Sackett, and Safronova}}]{leonard15a}
\bibinfo{author}{\bibfnamefont{R.~H.} \bibnamefont{Leonard}},
  \bibinfo{author}{\bibfnamefont{A.~J.} \bibnamefont{Fallon}},
  \bibinfo{author}{\bibfnamefont{C.~A.} \bibnamefont{Sackett}},
  \bibnamefont{and} \bibinfo{author}{\bibfnamefont{M.~S.}
  \bibnamefont{Safronova}}, \bibinfo{journal}{Phys. Rev. A}
  \textbf{\bibinfo{volume}{92}}, \bibinfo{pages}{052501}
  (\bibinfo{year}{2015}),
  \urlprefix\url{http://link.aps.org/doi/10.1103/PhysRevA.92.052501}.

\bibitem[{\citenamefont{Catani et~al.}(2009)\citenamefont{Catani, Barontini,
  Lamporesi, Rabatti, Thalhammer, Minardi, Stringari, and
  Inguscio}}]{catani09a}
\bibinfo{author}{\bibfnamefont{J.}~\bibnamefont{Catani}},
  \bibinfo{author}{\bibfnamefont{G.}~\bibnamefont{Barontini}},
  \bibinfo{author}{\bibfnamefont{G.}~\bibnamefont{Lamporesi}},
  \bibinfo{author}{\bibfnamefont{F.}~\bibnamefont{Rabatti}},
  \bibinfo{author}{\bibfnamefont{G.}~\bibnamefont{Thalhammer}},
  \bibinfo{author}{\bibfnamefont{F.}~\bibnamefont{Minardi}},
  \bibinfo{author}{\bibfnamefont{S.}~\bibnamefont{Stringari}},
  \bibnamefont{and} \bibinfo{author}{\bibfnamefont{M.}~\bibnamefont{Inguscio}},
  \bibinfo{journal}{Phys. Rev. Lett.} \textbf{\bibinfo{volume}{103}},
  \bibinfo{pages}{140401} (\bibinfo{year}{2009}),
  \urlprefix\url{http://link.aps.org/doi/10.1103/PhysRevLett.103.140401}.

\bibitem[{\citenamefont{Henson et~al.}(2015)\citenamefont{Henson, Khakimov,
  Dall, Baldwin, Tang, and Truscott}}]{henson15a}
\bibinfo{author}{\bibfnamefont{B.~M.} \bibnamefont{Henson}},
  \bibinfo{author}{\bibfnamefont{R.~I.} \bibnamefont{Khakimov}},
  \bibinfo{author}{\bibfnamefont{R.~G.} \bibnamefont{Dall}},
  \bibinfo{author}{\bibfnamefont{K.~G.~H.} \bibnamefont{Baldwin}},
  \bibinfo{author}{\bibfnamefont{L.-Y.} \bibnamefont{Tang}}, \bibnamefont{and}
  \bibinfo{author}{\bibfnamefont{A.~G.} \bibnamefont{Truscott}},
  \bibinfo{journal}{Phys. Rev. Lett.} \textbf{\bibinfo{volume}{115}},
  \bibinfo{pages}{043004} (\bibinfo{year}{2015}),
  \urlprefix\url{http://link.aps.org/doi/10.1103/PhysRevLett.115.043004}.

\bibitem[{\citenamefont{{Arora} et~al.}(2011)\citenamefont{{Arora},
  {Safronova}, and {Clark}}}]{arora11a}
\bibinfo{author}{\bibfnamefont{B.}~\bibnamefont{{Arora}}},
  \bibinfo{author}{\bibfnamefont{M.~S.} \bibnamefont{{Safronova}}},
  \bibnamefont{and} \bibinfo{author}{\bibfnamefont{C.~W.}
  \bibnamefont{{Clark}}}, \bibinfo{journal}{\pra}
  \textbf{\bibinfo{volume}{84}}, \bibinfo{eid}{043401} (\bibinfo{year}{2011}).

\bibitem[{\citenamefont{Lamporesi et~al.}(2010)\citenamefont{Lamporesi, Catani,
  Barontini, Nishida, Inguscio, and Minardi}}]{lamporesi10a}
\bibinfo{author}{\bibfnamefont{G.}~\bibnamefont{Lamporesi}},
  \bibinfo{author}{\bibfnamefont{J.}~\bibnamefont{Catani}},
  \bibinfo{author}{\bibfnamefont{G.}~\bibnamefont{Barontini}},
  \bibinfo{author}{\bibfnamefont{Y.}~\bibnamefont{Nishida}},
  \bibinfo{author}{\bibfnamefont{M.}~\bibnamefont{Inguscio}}, \bibnamefont{and}
  \bibinfo{author}{\bibfnamefont{F.}~\bibnamefont{Minardi}},
  \bibinfo{journal}{Phys. Rev. Lett.} \textbf{\bibinfo{volume}{104}},
  \bibinfo{pages}{153202} (\bibinfo{year}{2010}),
  \urlprefix\url{http://link.aps.org/doi/10.1103/PhysRevLett.104.153202}.

\bibitem[{\citenamefont{{Herold} et~al.}(2012)\citenamefont{{Herold}, {Vaidya},
  {Li}, {Rolston}, {Porto}, and {Safronova}}}]{herold12a}
\bibinfo{author}{\bibfnamefont{C.~D.} \bibnamefont{{Herold}}},
  \bibinfo{author}{\bibfnamefont{V.~D.} \bibnamefont{{Vaidya}}},
  \bibinfo{author}{\bibfnamefont{X.}~\bibnamefont{{Li}}},
  \bibinfo{author}{\bibfnamefont{S.~L.} \bibnamefont{{Rolston}}},
  \bibinfo{author}{\bibfnamefont{J.~V.} \bibnamefont{{Porto}}},
  \bibnamefont{and} \bibinfo{author}{\bibfnamefont{M.~S.}
  \bibnamefont{{Safronova}}}, \bibinfo{journal}{Phys.~Rev.~Lett.}
  \textbf{\bibinfo{volume}{109}}, \bibinfo{eid}{243003} (\bibinfo{year}{2012}).

\bibitem[{\citenamefont{Jiang et~al.}(2013)\citenamefont{Jiang, Tang, and
  Mitroy}}]{jiang13a}
\bibinfo{author}{\bibfnamefont{J.}~\bibnamefont{Jiang}},
  \bibinfo{author}{\bibfnamefont{L.~Y.} \bibnamefont{Tang}}, \bibnamefont{and}
  \bibinfo{author}{\bibfnamefont{J.}~\bibnamefont{Mitroy}},
  \bibinfo{journal}{Phys. Rev. A} \textbf{\bibinfo{volume}{87}},
  \bibinfo{pages}{032518} (\bibinfo{year}{2013}).

\bibitem[{\citenamefont{Jiang and Mitroy}(2013)}]{jiang13b}
\bibinfo{author}{\bibfnamefont{J.}~\bibnamefont{Jiang}} \bibnamefont{and}
  \bibinfo{author}{\bibfnamefont{J.}~\bibnamefont{Mitroy}},
  \bibinfo{journal}{Phys. Rev. A} \textbf{\bibinfo{volume}{88}},
  \bibinfo{pages}{032505} (\bibinfo{year}{2013}),
  \urlprefix\url{http://link.aps.org/doi/10.1103/PhysRevA.88.032505}.

\bibitem[{\citenamefont{{Grant} and {Quiney}}(2000)}]{grant00a}
\bibinfo{author}{\bibfnamefont{I.~P.} \bibnamefont{{Grant}}} \bibnamefont{and}
  \bibinfo{author}{\bibfnamefont{H.~M.} \bibnamefont{{Quiney}}},
  \bibinfo{journal}{\pra} \textbf{\bibinfo{volume}{62}}, \bibinfo{eid}{022508}
  (\bibinfo{year}{2000}).

\bibitem[{\citenamefont{Grant}(2007)}]{grant07a}
\bibinfo{author}{\bibfnamefont{I.~P.} \bibnamefont{Grant}},
  \emph{\bibinfo{title}{Relativistic Quantum Theory of Atoms and Molecules
  Theory and Computation}} (\bibinfo{publisher}{Springer},
  \bibinfo{address}{New York}, \bibinfo{year}{2007}).

\bibitem[{\citenamefont{Johnson et~al.}(1983)\citenamefont{Johnson, Kolb, and
  Huang}}]{johnson83a}
\bibinfo{author}{\bibfnamefont{W.~R.} \bibnamefont{Johnson}},
  \bibinfo{author}{\bibfnamefont{D.}~\bibnamefont{Kolb}}, \bibnamefont{and}
  \bibinfo{author}{\bibfnamefont{K.}~\bibnamefont{Huang}},
  \bibinfo{journal}{At.~Data~Nucl.~Data~Tables} \textbf{\bibinfo{volume}{28}},
  \bibinfo{pages}{333} (\bibinfo{year}{1983}).

\bibitem[{\citenamefont{Porsev and Derevianko}(2003)}]{porsev03a}
\bibinfo{author}{\bibfnamefont{S.~G.} \bibnamefont{Porsev}} \bibnamefont{and}
  \bibinfo{author}{\bibfnamefont{A.}~\bibnamefont{Derevianko}},
  \bibinfo{journal}{J.~Chem.~Phys.} \textbf{\bibinfo{volume}{119}},
  \bibinfo{pages}{844} (\bibinfo{year}{2003}).

\bibitem[{\citenamefont{Kramida et~al.}(2015)\citenamefont{Kramida,
  {Yu.~Ralchenko}, Reader, and {and NIST ASD Team}}}]{nistasd15a}
\bibinfo{author}{\bibfnamefont{A.}~\bibnamefont{Kramida}},
  \bibinfo{author}{\bibnamefont{{Yu.~Ralchenko}}},
  \bibinfo{author}{\bibfnamefont{J.}~\bibnamefont{Reader}}, \bibnamefont{and}
  \bibinfo{author}{\bibnamefont{{and NIST ASD Team}}},
  \bibinfo{howpublished}{{NIST Atomic Spectra Database (ver. 5.3), [Online].
  Available: {\tt{http://physics.nist.gov/asd}} [2016, June 15]. National
  Institute of Standards and Technology, Gaithersburg, MD.}}
  (\bibinfo{year}{2015}).

\bibitem[{\citenamefont{Safronova and Safronova}(2011)}]{safronova11aa}
\bibinfo{author}{\bibfnamefont{M.~S.} \bibnamefont{Safronova}}
  \bibnamefont{and} \bibinfo{author}{\bibfnamefont{U.~I.}
  \bibnamefont{Safronova}}, \bibinfo{journal}{Phys. Rev. A}
  \textbf{\bibinfo{volume}{83}}, \bibinfo{pages}{052508}
  (\bibinfo{year}{2011}),
  \urlprefix\url{http://link.aps.org/doi/10.1103/PhysRevA.83.052508}.

\bibitem[{\citenamefont{Safronova et~al.}(2004)\citenamefont{Safronova,
  Williams, and Clark}}]{safronova04a}
\bibinfo{author}{\bibfnamefont{M.~S.} \bibnamefont{Safronova}},
  \bibinfo{author}{\bibfnamefont{C.~J.} \bibnamefont{Williams}},
  \bibnamefont{and} \bibinfo{author}{\bibfnamefont{C.~W.} \bibnamefont{Clark}},
  \bibinfo{journal}{Phys.~Rev.~A} \textbf{\bibinfo{volume}{69}},
  \bibinfo{pages}{022509} (\bibinfo{year}{2004}).

\bibitem[{\citenamefont{Arora et~al.}(2012)\citenamefont{Arora, Nandy, and
  Sahoo}}]{arora12a}
\bibinfo{author}{\bibfnamefont{B.}~\bibnamefont{Arora}},
  \bibinfo{author}{\bibfnamefont{D.~K.} \bibnamefont{Nandy}}, \bibnamefont{and}
  \bibinfo{author}{\bibfnamefont{B.~K.} \bibnamefont{Sahoo}},
  \bibinfo{journal}{Phys. Rev. A} \textbf{\bibinfo{volume}{85}},
  \bibinfo{pages}{012506} (\bibinfo{year}{2012}).

\bibitem[{\citenamefont{{Pal} et~al.}(2007)\citenamefont{{Pal}, {Safronova},
  {Johnson}, {Derevianko}, and {Porsev}}}]{pal07a}
\bibinfo{author}{\bibfnamefont{R.}~\bibnamefont{{Pal}}},
  \bibinfo{author}{\bibfnamefont{M.~S.} \bibnamefont{{Safronova}}},
  \bibinfo{author}{\bibfnamefont{W.~R.} \bibnamefont{{Johnson}}},
  \bibinfo{author}{\bibfnamefont{A.}~\bibnamefont{{Derevianko}}},
  \bibnamefont{and} \bibinfo{author}{\bibfnamefont{S.~G.}
  \bibnamefont{{Porsev}}}, \bibinfo{journal}{Phys.~Rev.~A}
  \textbf{\bibinfo{volume}{75}}, \bibinfo{pages}{042515}
  (\bibinfo{year}{2007}).

\bibitem[{\citenamefont{Volz and Schmoranzer}(1996)}]{volz96a}
\bibinfo{author}{\bibfnamefont{U.}~\bibnamefont{Volz}} \bibnamefont{and}
  \bibinfo{author}{\bibfnamefont{H.}~\bibnamefont{Schmoranzer}},
  \bibinfo{journal}{Phys.~Scr.} \textbf{\bibinfo{volume}{T65}},
  \bibinfo{pages}{48} (\bibinfo{year}{1996}).

\bibitem[{\citenamefont{Simsarian et~al.}(1998)\citenamefont{Simsarian, Orozco,
  Sprouse, and Zhao}}]{simsarian98a}
\bibinfo{author}{\bibfnamefont{J.~E.} \bibnamefont{Simsarian}},
  \bibinfo{author}{\bibfnamefont{L.~A.} \bibnamefont{Orozco}},
  \bibinfo{author}{\bibfnamefont{G.~D.} \bibnamefont{Sprouse}},
  \bibnamefont{and} \bibinfo{author}{\bibfnamefont{W.~Z.} \bibnamefont{Zhao}},
  \bibinfo{journal}{Phys. Rev. A} \textbf{\bibinfo{volume}{57}},
  \bibinfo{pages}{2448} (\bibinfo{year}{1998}),
  \urlprefix\url{http://link.aps.org/doi/10.1103/PhysRevA.57.2448}.

\bibitem[{\citenamefont{Gutterres et~al.}(2002)\citenamefont{Gutterres, Amiot,
  Fioretti, Gabbanini, Mazzoni, and Dulieu}}]{gutterres02a}
\bibinfo{author}{\bibfnamefont{R.~F.} \bibnamefont{Gutterres}},
  \bibinfo{author}{\bibfnamefont{C.}~\bibnamefont{Amiot}},
  \bibinfo{author}{\bibfnamefont{A.}~\bibnamefont{Fioretti}},
  \bibinfo{author}{\bibfnamefont{C.}~\bibnamefont{Gabbanini}},
  \bibinfo{author}{\bibfnamefont{M.}~\bibnamefont{Mazzoni}}, \bibnamefont{and}
  \bibinfo{author}{\bibfnamefont{O.}~\bibnamefont{Dulieu}},
  \bibinfo{journal}{Phys. Rev. A} \textbf{\bibinfo{volume}{66}},
  \bibinfo{pages}{024502} (\bibinfo{year}{2002}),
  \urlprefix\url{http://link.aps.org/doi/10.1103/PhysRevA.66.024502}.

\bibitem[{\citenamefont{Mitroy et~al.}(2010)\citenamefont{Mitroy, Safronova,
  and Clark}}]{mitroy10a}
\bibinfo{author}{\bibfnamefont{J.}~\bibnamefont{Mitroy}},
  \bibinfo{author}{\bibfnamefont{M.~S.} \bibnamefont{Safronova}},
  \bibnamefont{and} \bibinfo{author}{\bibfnamefont{C.~W.} \bibnamefont{Clark}},
  \bibinfo{journal}{J.~Phys.~B} \textbf{\bibinfo{volume}{43}},
  \bibinfo{pages}{202001} (\bibinfo{year}{2010}).

\bibitem[{\citenamefont{Lee et~al.}(1975)\citenamefont{Lee, Das, and
  Sternheimer}}]{lee75a}
\bibinfo{author}{\bibfnamefont{T.}~\bibnamefont{Lee}},
  \bibinfo{author}{\bibfnamefont{T.~P.} \bibnamefont{Das}}, \bibnamefont{and}
  \bibinfo{author}{\bibfnamefont{R.~M.} \bibnamefont{Sternheimer}},
  \bibinfo{journal}{Phys. Rev. A} \textbf{\bibinfo{volume}{11}},
  \bibinfo{pages}{1784} (\bibinfo{year}{1975}).

\bibitem[{\citenamefont{Snider}(1966)}]{snider66a}
\bibinfo{author}{\bibfnamefont{J.~L.} \bibnamefont{Snider}},
  \bibinfo{journal}{Phys.~Lett.~} \textbf{\bibinfo{volume}{21}},
  \bibinfo{pages}{172} (\bibinfo{year}{1966}).

\bibitem[{\citenamefont{{Migdalek} and {Kim}}(1998)}]{migdalek98a}
\bibinfo{author}{\bibfnamefont{J.}~\bibnamefont{{Migdalek}}} \bibnamefont{and}
  \bibinfo{author}{\bibfnamefont{Y.-K.} \bibnamefont{{Kim}}},
  \bibinfo{journal}{J.~Phys.~B} \textbf{\bibinfo{volume}{31}},
  \bibinfo{pages}{1947} (\bibinfo{year}{1998}).

\bibitem[{\citenamefont{Tang Yong-Bo}(2014)}]{tang14a}
\bibinfo{author}{\bibfnamefont{Q.~H.-X.} \bibnamefont{Tang Yong-Bo},
  \bibfnamefont{Li~Cheng-Bin}}, \bibinfo{journal}{Chinese Physics B}
  \textbf{\bibinfo{volume}{23}}, \bibinfo{eid}{63101}
  (pages~\bibinfo{numpages}{0}) (\bibinfo{year}{2014}),
  \urlprefix\url{http://cpb.iphy.ac.cn/EN/abstract/article_59819.shtml}.

\bibitem[{\citenamefont{Mitroy and Novikov}(2003)}]{mitroy03a}
\bibinfo{author}{\bibfnamefont{J.}~\bibnamefont{Mitroy}} \bibnamefont{and}
  \bibinfo{author}{\bibfnamefont{S.}~\bibnamefont{Novikov}},
  \bibinfo{journal}{Phys.~Rev.~Lett.} \textbf{\bibinfo{volume}{90}},
  \bibinfo{pages}{183202} (\bibinfo{year}{2003}).

\bibitem[{\citenamefont{{Lim} et~al.}(2005)\citenamefont{{Lim},
  {Schwerdtfeger}, {Metz}, and {Stoll}}}]{lim05a}
\bibinfo{author}{\bibfnamefont{I.~S.} \bibnamefont{{Lim}}},
  \bibinfo{author}{\bibfnamefont{P.}~\bibnamefont{{Schwerdtfeger}}},
  \bibinfo{author}{\bibfnamefont{B.}~\bibnamefont{{Metz}}}, \bibnamefont{and}
  \bibinfo{author}{\bibfnamefont{H.}~\bibnamefont{{Stoll}}},
  \bibinfo{journal}{J.~Chem.~Phys.} \textbf{\bibinfo{volume}{122}},
  \bibinfo{pages}{104103} (\bibinfo{year}{2005}).

\bibitem[{\citenamefont{Kaur et~al.}(2014)\citenamefont{Kaur, Kaur, Arora, and
  Sahoo}}]{kaur14a}
\bibinfo{author}{\bibfnamefont{K.}~\bibnamefont{Kaur}},
  \bibinfo{author}{\bibfnamefont{J.}~\bibnamefont{Kaur}},
  \bibinfo{author}{\bibfnamefont{B.}~\bibnamefont{Arora}}, \bibnamefont{and}
  \bibinfo{author}{\bibfnamefont{B.~K.} \bibnamefont{Sahoo}},
  \bibinfo{journal}{Phys. Rev. B} \textbf{\bibinfo{volume}{90}},
  \bibinfo{pages}{245405} (\bibinfo{year}{2014}),
  \urlprefix\url{http://link.aps.org/doi/10.1103/PhysRevB.90.245405}.

\bibitem[{\citenamefont{Kaur et~al.}(2015)\citenamefont{Kaur, Nandy, Arora, and
  Sahoo}}]{kaur15a}
\bibinfo{author}{\bibfnamefont{J.}~\bibnamefont{Kaur}},
  \bibinfo{author}{\bibfnamefont{D.~K.} \bibnamefont{Nandy}},
  \bibinfo{author}{\bibfnamefont{B.}~\bibnamefont{Arora}}, \bibnamefont{and}
  \bibinfo{author}{\bibfnamefont{B.~K.} \bibnamefont{Sahoo}},
  \bibinfo{journal}{Phys. Rev. A} \textbf{\bibinfo{volume}{91}},
  \bibinfo{pages}{012705} (\bibinfo{year}{2015}),
  \urlprefix\url{http://link.aps.org/doi/10.1103/PhysRevA.91.012705}.

\bibitem[{\citenamefont{Safronova et~al.}(1999)\citenamefont{Safronova,
  Johnson, and Derevianko}}]{safronova99a}
\bibinfo{author}{\bibfnamefont{M.~S.} \bibnamefont{Safronova}},
  \bibinfo{author}{\bibfnamefont{W.~R.} \bibnamefont{Johnson}},
  \bibnamefont{and}
  \bibinfo{author}{\bibfnamefont{A.}~\bibnamefont{Derevianko}},
  \bibinfo{journal}{Phys.~Rev.~A} \textbf{\bibinfo{volume}{60}},
  \bibinfo{pages}{4476} (\bibinfo{year}{1999}).

\bibitem[{\citenamefont{Derevianko et~al.}(1999)\citenamefont{Derevianko,
  Johnson, Safronova, and Babb}}]{derevianko99a}
\bibinfo{author}{\bibfnamefont{A.}~\bibnamefont{Derevianko}},
  \bibinfo{author}{\bibfnamefont{W.~R.} \bibnamefont{Johnson}},
  \bibinfo{author}{\bibfnamefont{M.~S.} \bibnamefont{Safronova}},
  \bibnamefont{and} \bibinfo{author}{\bibfnamefont{J.~F.} \bibnamefont{Babb}},
  \bibinfo{journal}{Phys.~Rev.~Lett.} \textbf{\bibinfo{volume}{82}},
  \bibinfo{pages}{3589} (\bibinfo{year}{1999}).

\bibitem[{\citenamefont{Molof et~al.}(1974)\citenamefont{Molof, Schwartz,
  Miller, and Bederson}}]{molof74a}
\bibinfo{author}{\bibfnamefont{R.~W.} \bibnamefont{Molof}},
  \bibinfo{author}{\bibfnamefont{H.~L.} \bibnamefont{Schwartz}},
  \bibinfo{author}{\bibfnamefont{T.~M.} \bibnamefont{Miller}},
  \bibnamefont{and} \bibinfo{author}{\bibfnamefont{B.}~\bibnamefont{Bederson}},
  \bibinfo{journal}{Phys.~Rev.~A} \textbf{\bibinfo{volume}{10}},
  \bibinfo{pages}{1131} (\bibinfo{year}{1974}).

\bibitem[{\citenamefont{Holmgren et~al.}(2010)\citenamefont{Holmgren, Revelle,
  Lonij, and Cronin}}]{holmgren10a}
\bibinfo{author}{\bibfnamefont{W.~F.} \bibnamefont{Holmgren}},
  \bibinfo{author}{\bibfnamefont{M.~C.} \bibnamefont{Revelle}},
  \bibinfo{author}{\bibfnamefont{V.~P.~A.} \bibnamefont{Lonij}},
  \bibnamefont{and} \bibinfo{author}{\bibfnamefont{A.~D.}
  \bibnamefont{Cronin}}, \bibinfo{journal}{Phys.~Rev.~A}
  \textbf{\bibinfo{volume}{81}}, \bibinfo{pages}{053607}
  (\bibinfo{year}{2010}).

\bibitem[{\citenamefont{Gregoire et~al.}(2015)\citenamefont{Gregoire, Hromada,
  Holmgren, Trubko, and Cronin}}]{gregoire15a}
\bibinfo{author}{\bibfnamefont{M.~D.} \bibnamefont{Gregoire}},
  \bibinfo{author}{\bibfnamefont{I.}~\bibnamefont{Hromada}},
  \bibinfo{author}{\bibfnamefont{W.~F.} \bibnamefont{Holmgren}},
  \bibinfo{author}{\bibfnamefont{R.}~\bibnamefont{Trubko}}, \bibnamefont{and}
  \bibinfo{author}{\bibfnamefont{A.~D.} \bibnamefont{Cronin}},
  \bibinfo{journal}{Phys. Rev. A} \textbf{\bibinfo{volume}{92}},
  \bibinfo{pages}{052513} (\bibinfo{year}{2015}),
  \urlprefix\url{http://link.aps.org/doi/10.1103/PhysRevA.92.052513}.

\bibitem[{\citenamefont{Arimondo et~al.}(1977)\citenamefont{Arimondo, Inguscio,
  and Violino}}]{arimondo77a}
\bibinfo{author}{\bibfnamefont{E.}~\bibnamefont{Arimondo}},
  \bibinfo{author}{\bibfnamefont{M.}~\bibnamefont{Inguscio}}, \bibnamefont{and}
  \bibinfo{author}{\bibfnamefont{P.}~\bibnamefont{Violino}},
  \bibinfo{journal}{Rev. Mod. Phys.} \textbf{\bibinfo{volume}{49}},
  \bibinfo{pages}{31} (\bibinfo{year}{1977}).

\bibitem[{\citenamefont{Armstrong}(1971)}]{armstrong71a}
\bibinfo{author}{\bibfnamefont{L.~J.} \bibnamefont{Armstrong}},
  \emph{\bibinfo{title}{Theory of the Hyperfine Structure of Free Atoms}}
  (\bibinfo{publisher}{Wiley-Interscience}, \bibinfo{address}{New York},
  \bibinfo{year}{1971}).

\bibitem[{\citenamefont{{Rapol} et~al.}(2003)\citenamefont{{Rapol}, {Krishna},
  and {Natarajan}}}]{rapol03a}
\bibinfo{author}{\bibfnamefont{U.~D.} \bibnamefont{{Rapol}}},
  \bibinfo{author}{\bibfnamefont{A.}~\bibnamefont{{Krishna}}},
  \bibnamefont{and}
  \bibinfo{author}{\bibfnamefont{V.}~\bibnamefont{{Natarajan}}},
  \bibinfo{journal}{European Physical Journal D} \textbf{\bibinfo{volume}{23}},
  \bibinfo{pages}{185} (\bibinfo{year}{2003}), \eprint{physics/0307090}.

\bibitem[{\citenamefont{Steck}(2010)}]{steck10a}
\bibinfo{author}{\bibfnamefont{D.~A.} \bibnamefont{Steck}},
  \bibinfo{journal}{available online at
  http://steck.us/alkalidata(revision2.1.4)}  (\bibinfo{year}{2010}).

\bibitem[{\citenamefont{{Angstmann} et~al.}(2006)\citenamefont{{Angstmann},
  {Dzuba}, and {Flambaum}}}]{angstmann06a}
\bibinfo{author}{\bibfnamefont{E.~J.} \bibnamefont{{Angstmann}}},
  \bibinfo{author}{\bibfnamefont{V.~A.} \bibnamefont{{Dzuba}}},
  \bibnamefont{and} \bibinfo{author}{\bibfnamefont{V.~V.}
  \bibnamefont{{Flambaum}}}, \bibinfo{journal}{Phys.~Rev.~Lett.}
  \textbf{\bibinfo{volume}{97}}, \bibinfo{pages}{040802}
  (\bibinfo{year}{2006}).

\bibitem[{\citenamefont{{Safronova} et~al.}(2010)\citenamefont{{Safronova},
  Jiang, Arora, Clark, Kozlov, {Safronova}, and Johnson}}]{safronova10a}
\bibinfo{author}{\bibfnamefont{M.~S.} \bibnamefont{{Safronova}}},
  \bibinfo{author}{\bibfnamefont{D.}~\bibnamefont{Jiang}},
  \bibinfo{author}{\bibfnamefont{B.}~\bibnamefont{Arora}},
  \bibinfo{author}{\bibfnamefont{C.~W.} \bibnamefont{Clark}},
  \bibinfo{author}{\bibfnamefont{M.~G.} \bibnamefont{Kozlov}},
  \bibinfo{author}{\bibfnamefont{U.~I.} \bibnamefont{{Safronova}}},
  \bibnamefont{and} \bibinfo{author}{\bibfnamefont{W.~R.}
  \bibnamefont{Johnson}}, \bibinfo{journal}{IEEE Trans. Ultrason.
  Ferroelectrics and Frequency Control} \textbf{\bibinfo{volume}{57}},
  \bibinfo{pages}{94} (\bibinfo{year}{2010}).

\bibitem[{\citenamefont{Mowat}(1972)}]{mowat72a}
\bibinfo{author}{\bibfnamefont{J.~R.} \bibnamefont{Mowat}},
  \bibinfo{journal}{Phys. Rev. A} \textbf{\bibinfo{volume}{5}},
  \bibinfo{pages}{1059} (\bibinfo{year}{1972}).

\bibitem[{\citenamefont{Dallal and Ozeri}(2015)}]{dallal15a}
\bibinfo{author}{\bibfnamefont{Y.}~\bibnamefont{Dallal}} \bibnamefont{and}
  \bibinfo{author}{\bibfnamefont{R.}~\bibnamefont{Ozeri}},
  \bibinfo{journal}{Phys. Rev. Lett.} \textbf{\bibinfo{volume}{115}},
  \bibinfo{pages}{183001} (\bibinfo{year}{2015}),
  \urlprefix\url{http://link.aps.org/doi/10.1103/PhysRevLett.115.183001}.

\end{thebibliography}
\end{document}